\pdfoutput=1
\documentclass{aa}
\usepackage{pifont}
\usepackage{amsmath}
\usepackage{graphicx}
\usepackage{txfonts}
\usepackage{url}
\usepackage{subfigure}
\usepackage{longtable}
\usepackage{lscape}
\usepackage{natbib}
\usepackage[colorlinks,linkcolor=red,anchorcolor=green,citecolor=blue]{hyperref}
\usepackage[normalem]{ulem}

\newcommand{\msol}{\mbox{M$_\odot$}}
\newcommand{\rsol}{\mbox{R$_\odot$}}

\newcommand{\mjyb}{{\rm mJy\,beam}^{-1}}
\newcommand{\mjyp}{{\rm mJy\,pixel}^{-1}}

\newcommand{\mujyp}{{\rm {\mu}Jy\,pixel}^{-1}}

\newcommand{\dd}{\, {\rm d}}	    
\newcommand{\env}{{\rm env}}	    
\newcommand{\disk}{{\rm disk}}	    
\newcommand{\rcyl}{r_{\rm cyl}}	    
\newcommand{\K}{\rm K}	            
\newcommand{\nm}{\, \rm nm}	    
\newcommand{\cm}{\, \rm cm}	    
\newcommand{\mum}{\, \rm \mu m}	    
\newcommand{\lsun}{L_\odot}	    
\newcommand{\au}{\rm au}	    

\newcommand{\Ka}{\mbox{K$_{\rm a}$}}

\begin{document}

    \title{Pebbles in an Embedded Protostellar Disk: The Case of CB\,26}
    \authorrunning{C.-P. Zhang et al.}
    \titlerunning{The Bok globule CB\,26}

    \author{
    Chuan-Peng Zhang\inst{1,2}
    \and
    Ralf Launhardt\inst{2}
    \and
    Yao Liu\inst{3}
    \and
    John J. Tobin\inst{4}
    \and
    Thomas Henning\inst{2}
    }

    \institute{
    National Astronomical Observatories, Chinese Academy of Sciences, 100101 Beijing, P.R. China\\
    \email{cpzhang@nao.cas.cn}
    \and
    Max-Planck-Institut f\"ur Astronomie, K\"onigstuhl 17, D-69117 Heidelberg, Germany \\
    \email{rl@mpia.de}
    \and
    Purple Mountain Observatory, Chinese Academy of Sciences, 2 West Beijing Road, 210008 Nanjing, P.R. China
    \and
    National Radio Astronomy Observatory, 520 Edgemont Road, Charlottesville, VA 22903, USA 
    } 



   \abstract
    {Planetary cores are thought to form in proto-planetary disks via the growth of dusty solid material. However, it is unclear how early this process begins.}
   {We study the physical structure and grain growth in the edge-on disk that surrounds the $\approx$1\,Myr old low-mass ($\approx$0.55\,\msol) protostar embedded in the Bok Globule CB26 to examine how much grain growth has already occurred in the protostellar phase.}
   {We combine the SED between 0.9$\mum$ and 6.4\,cm with high angular resolution continuum maps at 1.3, 2.9, and 8.1\,mm, and use the radiative transfer code \texttt{RADMC-3D} to conduct a detailed modelling of the dust emission from the disk and envelope of CB\,26.}
   {Given the presence of central disk cavity, we infer inner and outer disk radii of $16^{+37}_{-8}$\,au and $172\pm22$\,au, respectively. The total gas mass in the disk is $7.6\times10^{-2}$\,\msol, which amounts to $\approx$14\% of the mass of the central star. The inner disk contains a compact free-free emission region, which could be related to either a jet or a photoevaporation region. The thermal dust emission from the outer disk is optically thin at mm wavelengths, while the emission from the inner disk midplane is moderately optically thick. Our best-fit radiative transfer models indicate that the dust grains in the disk have already grown to pebbles with diameters of the order of 10\,cm in size. Residual 8.1\,mm emission suggests the presence of even larger particles in the inner disk. For the optically thin mm dust emission from the outer disk, we derive a mean opacity slope of \mbox{$\beta_{\rm mm}\approx0.7\pm0.4$}, which is consistent with the presence of large dust grains.}
   {The presence of cm-sized bodies in the CB\,26 disk indicates that solids grow rapidly already during the first million years in a protostellar disk. It is thus possible that Class\,II disks are already seeded with large particles and may contain even planetesimals.}

   \keywords{(stars:) circumstellar matter -- protoplanetary disks -- radiative transfer -- stars: individual: CB\,26}

   \maketitle
%

\section{Introduction}    
\label{sect_intro}

The formation of a circumstellar disk is a key step in the process of forming a planetary system \citep[e.g.,][]{Williams2011,Matthews2014,Tobin2012,Tobin2015,Bayo2019,Ubeira2019}. A necessary prerequisite to understand the formation of a planetary system is the comprehension of when the dust grains start to grow and how the particle size distribution evolves with time. Protostellar disks radiate most strongly in the infrared, dominated by thermal dust emission and scattered stellar light from dust in the circumstellar envelope and the disk's upper optically thin layers \citep{Williams2011,Williams2016}. However, while the infrared emission is usually optically thick, the emission at millimeter wavelengths is generally optically thin and therefore provides a more precise way to diagnose the dust grain distribution and the physical structure of the disk regime through high-angular-resolution interferometric observations. Therefore, high-angular resolution millimeter observations coupled with radiative transfer modelling enable the physical structure and dust grain size distributions to be examined for disks, especially the dust grain growth along with the evolution.

CB\,26 (L1439) is a small cometary-shaped Bok globule located at $\sim$10$^\circ$ north of the Taurus-Auriga dark cloud at a distance of $140\pm20$\,pc \citep{Launhardt2001, Launhardt2010, Launhardt2013}. The embedded infrared (IR) source IRAS\,04559+5200 is associated with a Class\,I young stellar object \citep[YSO;][]{Launhardt2001,Stecklum2004}. Single-dish mm continuum observations revealed that IRAS\,04559+5200 is associated with an unresolved 1.3\,mm continuum source \citep{Launhardt1997}. \citet{Henning2001} then found that an optically thin asymmetric envelope with a well-ordered magnetic field directed along P.A.\,$\approx25^{\circ}$ is surrounding the source. Follow-up interferometric observations at millimeter wavelengths by \citet{Launhardt2001} resolved the central continuum source and showed that it is a young edge-on circumstellar disk with a radius of $\sim$200\,au and a mass of $\sim$0.1\,\msol. 

Later, \citet{Launhardt2009} detected a collimated bipolar molecular outflow of $\sim$2000\,au in total length, escaping perpendicular to the plane of the well-resolved circumstellar disk. The outflow is launched from the disk and has a Keplerian rotation profile that reaches out to at least 1000\,au above the disk. A total dynamical mass of the central star(s) of \mbox{$M_{\star} = 0.55\pm 0.1$\,\msol} was derived by fitting a Keplerian disk model to the $^{12}$CO $J=2-1$ velocity structure of the disk \citep{Launhardt2020}. Based on radiative transfer models, \citet{Launhardt2009}, \citet{Sauter2009}, and \citet{Akimkin2012} consistently concluded that the dust disk has an optically thin inner hole with radius \mbox{$r_{\rm in}\approx30-50$\,au}. The age of the system was estimated by \citet{Launhardt2009} to be $1^{+1}_{-0.5}$\,Myr. Hence, CB\,26 can be considered as a still partially embedded disk/jet/outflow system at the transition from a protostellar accretion disk to a protoplanetary transition disk \citep[e.g.,][]{espaillat2014}.

\citet{Sauter2009} and \citet{Akimkin2012} have successfully modelled the disk of CB\,26 using different radiative transfer codes. However, their models still have some shortcomings. \citet{Sauter2009} derive a millimeter wavelength opacity slope between 1.1 and 2.7\,mm of $\beta_{\rm mm}=1.1\pm0.27$, assuming optically thin emission from both the disk and envelope. This would be indicative that at least millimeter-size dust grains should be present according to the relation derived by \citet{Ricci2010b}.  Yet, their best-fit maximum dust grain size in the disk is only \mbox{$a_{\rm max}^{\rm disk} = 2.5\mum$}.
\citet{Akimkin2012}, on the other hand, neither considered the envelope nor the overall spectral energy distribution (SED) in their modelling. They only derive an upper limit on the maximum dust grain size of $a_{\rm max}^{\rm disk} < 200\mum$. Their best-fit modelled inclination angle of the disk is 78$^{\circ}$, which is 7$^{\circ}$ lower than the 85$^{\circ}$ derived by \citet{Launhardt2009} and \citet{Sauter2009}. They also state that their models have a wide range of degeneracy between the thermal and density parameters of the disk. Therefore, in the presence of new high-quality observational data with higher angular resolution and covering longer wavelengths, we carry out a new study of this source.

We re-compile the SED and include new observational data at higher angular resolution at 1.3 and 2.9\,mm and additional longer-wavelength data than in previous studies of CB\,26. This allows us to more precisely model the dust grain distribution and the physical structure of the disk and envelope. In Section\,\ref{sect_obser}, we describe our high angular resolution millimeter observations. Section\,\ref{sect_model_modelling} introduces our employed models and the modelling process. In Section\,\ref{sect_results}, we present our modelling results. In Section\,\ref{sect_discussion}, we discuss the free-free emission, inner hole, grain growth, and Toomre stability of the disk. Section\,\ref{sect_summary} summarizes this study.


\section{Observations and data reduction}
\label{sect_obser}

\begin{table}
  \caption{Photometric data points for CB\,26.}\label{tab_photometric}
  \centering \small
  \begin{tabular}{lcccc}
    \hline \hline
      $\lambda$ & Flux & Aperture & Instrument & Ref. \\
      $\mu$m & mJy & $^{\prime\prime}$ &  & \\
    \hline 
   $0.90 $ &  $0.062  \pm 0.019 $ & $24  $ & CAHA 3.5m         & (1) \\
   $1.25 $ &  $2.2    \pm 0.2   $ & $12  $ & CAHA 3.5m         & (1) \\ 
   $1.65 $ &  $8.2    \pm 0.8   $ & $12  $ & CAHA 3.5m         & (1) \\ 
   $2.20 $ &  $17.1   \pm 1.7   $ & $12  $ & CAHA 3.5m         & (1) \\
   $3.6  $ &  $18.3   \pm 0.8   $ & $12  $ & \textit{Spitzer} IRAC1     & (2) \\
   $4.5  $ &  $17.0   \pm 0.8   $ & $12  $ & \textit{Spitzer} IRAC2     & (2) \\
   $4.6  $ &  $17.96  \pm 0.5   $ & $12  $ & WISE band 2       & (3) \\
   $5.8  $ &  $12.5   \pm 0.7   $ & $12  $ & \textit{Spitzer} IRAC3     & (2) \\
   $8.0  $ &  $6.8    \pm 0.5   $ & $12  $ & \textit{Spitzer} IRAC4     & (2) \\
   $11.6 $ &  $4.5    \pm 0.5   $ & $8   $ & WISE band 3       & (3) \\
   $24   $ &  $160.6  \pm 16.0  $ & $30  $ & \textit{Spitzer} MIPS1     & (2) \\
   $65   $ &  $4398   \pm 274   $ & $9 0 $ & \textit{Akari}             & (3) \\
   $70   $ &  $5555   \pm 1000  $ & $120 $ & \textit{Spitzer} MIPS2     & (2) \\
   $100  $ &  $11000  \pm 3000  $ & $125 $ & \textit{Herschel} PACS100  & (3) \\ 
   $160  $ &  $19000  \pm 4000  $ & $80  $ & \textit{Herschel} PACS160  & (3) \\
   $250  $ &  $23700  \pm 5000  $ & $ 90 $ &  \textit{Herschel} SPIRE   & (3) \\
   $350  $ &  $15000  \pm 3000  $ & $ 90 $ &  \textit{Herschel} SPIRE   & (3) \\
   $450  $ &  $6700   \pm 2500  $ & $ 60 $ &  SCUBA            & (3) \\
   $500  $ &  $6200   \pm 2000  $ & $ 90 $ &  \textit{Herschel} SPIRE   & (3) \\
   $850  $ &  $1000   \pm 300   $ & $ 60 $ &  SCUBA            & (3) \\
   $1110 $ &  $238    \pm 45    $ & $ 5  $ &  SMA              & (4) \\                   
   $1300 $ &  $120    \pm 30    $ & $ 5  $ &  OVRO+PdPI        & (4) \\         
   $2900 $ &  $14.5   \pm 3.0   $ & $ 5  $ &  CARMA            & (4) \\                   
   $8100 $ &  $1.10   \pm 0.15  $ & $ 5  $ &  VLA A+B          & (4) \\                    
   $10300$ &  $0.65   \pm 0.1   $ & $ 5  $ &  VLA A+B          & (4) \\                    
   $40000$ &  $0.050  \pm 0.01  $ & $ 5  $ &  VLA B            & (4) \\
   $64000$ &  $0.047  \pm 0.02  $ & $ 5  $ &  VLA A+B          & (4) \\
    \hline
  \end{tabular}
  \tablebib{
  (1)~\mbox{\citet{Stecklum2004}};
  (2)~\mbox{\citet{Sauter2009}};
  (3)~\mbox{\citet{Launhardt2013}};
  (4)~This work.}

\end{table}

\begin{table}
  \caption{Beam sizes and noise levels of the millimeter-maps}
  \label{tab_beam}
  \centering
  \begin{tabular}{cccc}
    \hline \hline
      Instrument & $\lambda$ & Synth. HPBW; P.A. & 1$\sigma$ rms\\
      & mm & $\arcsec\times\arcsec$; \,\,\,\,$^\circ$ & $\mjyb$ \\
    \hline
      SMA &        $1.1$ & $1.00\times0.84$; $-88.9$  & 2.6 \\
      OVRO+PdBI &  $1.3$ & $0.39\times0.36$;  43.3  & 0.30 \\
      CARMA     &  $2.9$ & $0.72\times0.51$;  82.2  & 0.25 \\
      VLA  &       $8.1$ & $0.133\times0.126$; $-66.0$  & 0.0071 \\
    \hline
  \end{tabular}
\end{table}

\begin{figure}[htp]
\centering
\includegraphics[width=0.48\textwidth, angle=0]{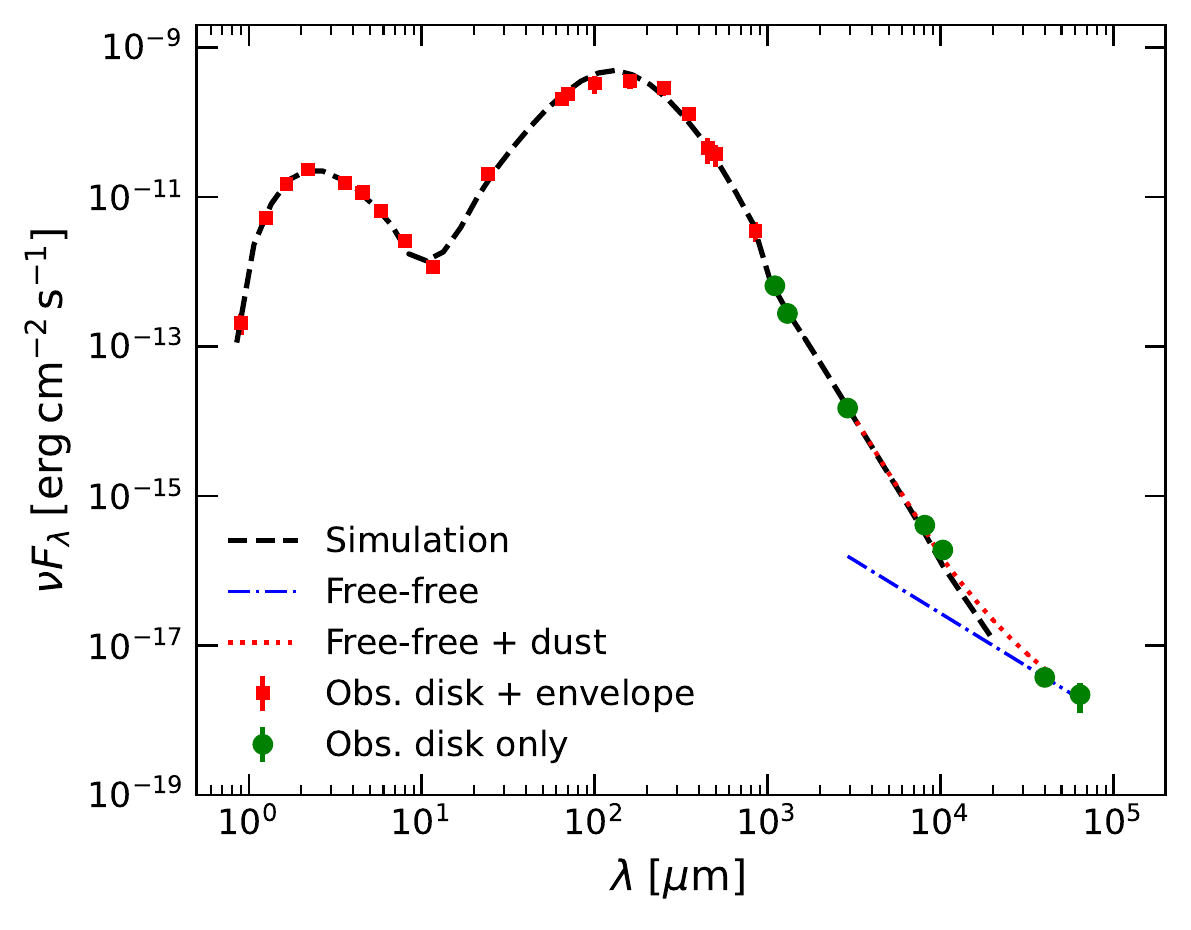}
\caption{Spectral energy distribution (SED) of CB\,26. The red squares and green dots with error bars represent the observational data (Table\,\ref{tab_photometric}). The black dashed line shows our best-fit model SED. The blue dashed-dotted line represents the adopted contribution from free-free emission. The red dotted line corresponds to the sum flux of dust and free-free emission between 2.9\,mm and 4.0\,cm. The jump in the SED at $\sim$10$^3\mum$ is due to the transition from total flux measurements at shorter wavelengths to interferometric data at longer wavelengths, which do not recover the total flux from the extended envelope emission.}
\label{fig_sed}
\end{figure}

The YSO in CB\,26 has been observed at many wavelengths, ranging from $0.9\mum$ to 6.4\,cm. Table\,\ref{tab_photometric} summarizes the photometric data used in this study, and Fig.\,\ref{fig_sed} shows the spectral energy distribution (SED) of CB\,26 along with our best-fit model SED (Sect.\,\ref{sect_fit}). Most of the observations and data reduction at wavelengths shorter than 1\,mm have been presented in previous papers (see references in Table\,\ref{tab_photometric}) and are not further described here. In this section, we only present the high angular resolution millimeter observations at 1.1, 1.3, 2.9, 8.1, 10.3, 40 and 64\,mm, obtained with the Sub-millimeter Array (SMA), Owens Valley Radio Observatory (OVRO), Plateau de Bure Interferometer (PdBI), Combined Array for Research in Millimeter Astronomy (CARMA), and the Karl G. Jansky Very Large Array (VLA) instruments.


\subsection{SMA}
\label{sect_obs_sma}

In December 2006, the SMA \citep{Ho2004} was used to observe CB\,26 at 267\,--\,277\,GHz ($\sim$1.1\,mm) with the extended and compact configurations covering a baseline range of 10\,--\,173\,k$\lambda$. Typical system temperatures were 350\,--\,500\,K during the observations. The raw data were edited and calibrated using the IDL MIR package\footnote{\url{https://www.cfa.harvard.edu/sma/smaMIR/}}. The amplitude and phase gain calibrators were the quasars B0355+508 and 3C111, and the bandpass calibrator was the quasar 3C279. Uranus was used for absolute flux calibration, resulting in a relative uncertainty of 20\%\,--\,30\%. Using line-free channels in both lower and upper sidebands, robust $uv$-weighting (weighting parameter 0) was used to construct the 1.1\,mm continuum map. The image data cubes were then exported to GREG in the GILDAS package\footnote{\url{http://www.iram.fr/IRAMFR/GILDAS/}} for further continuum map analysis. Effective synthesised beam size is listed in Table\,\ref{tab_beam}.


\subsection{OVRO}

CB\,26 was also observed with the Owens Valley Radio Observatory (OVRO) millimeter array between January 2000 and December 2001. Four configurations of the six 10.4\,m antennas provided a $uv$ baseline range of 12\,--\,400\,k$\lambda$\ at 1.3\,mm (232\,GHz). Average single sideband (SSB) system temperatures of the SIS receivers were 300\,--\,600\,K at 232\,GHz. The quasar (QSO) 0355+508, which is located 9$^\circ$\ away from CB\,26, served as amplitude and phase gain calibrator. The bandpass was calibrated on QSO 3C273. The absolute flux scale was calibrated using the planets Uranus and Neptune. The raw data were exported to the MMA software package \citep{Scoville1993} for editing and calibration. The MIRIAD software package \citep{Sault1995} was used for cleaning, mapping and data analysis. The OVRO observations and data are described in more detail in \citet{Launhardt2001}. For this study, the OVRO data were not used separately, but were combined with the PdBI data to obtain a 1.3\,mm map with the best possible {\it uv} coverage and signal-to-noise ratio (see below).


\subsection{PdBI}

In November and December 2005, IRAM\footnote{IRAM is supported by INSU/CNRS (France), MPG (Germany) and IGN (Spain).} PdBI observations of CB\,26 were carried out using D configuration with 5 antennas and C configuration with 6 antennas, respectively. The $uv$ baselines of the two array configurations range from 16 to 175\,m. The phase center was at $\rm \alpha_{J2000} = 04^h59^m50.74^s$, $\rm \delta_{J2000} = 52^{\circ}04'43.80''$. Two receivers were used simultaneously and tuned SSB to around 89\,GHz (3.4\,mm) and 230\,GHz (1.3\,mm), respectively. The 3.4\,mm data are not used here because of the superior quality of the CARMA 3\,mm data.
The QSO 0355+508 was used to determine the time-dependent complex antenna gains. The correlator bandpass was calibrated on 3C454.3 and 3C273, and the absolute flux density scale was derived from observations of MWC\,349. The flux calibration uncertainty is estimated to be $\la$20\%. The primary beam size at 1.3\,mm was 22$''$. The SSB system temperature at 230\,GHz was in the range 250\,--\,450\,K. The data were calibrated and imaged using the GILDAS software. 

For imaging the continuum emission, we combined the calibrated continuum $uv$ tables from OVRO and PdBI after applying a $uv$ shift of $\approx$0.2$''$ to the OVRO data. This was necessary to account for a small source center shift which is likely related to a combination of the unknown proper motion of CB\,26 and uncertainties in calibrator coordinates and phase calibration errors. The 1.3\,mm data were obtained at the same frequency. Imaging and deconvolution of the continuum data was done with robust $uv$-weighting (weighting parameter 0), resulting in the synthesized beam sizes and map noise levels listed in Table\,\ref{tab_beam}.


\subsection{CARMA}

The CARMA was a 23-element interferometer located in the Inyo Mountains of California, near Owens Valley. The main CARMA array was used for these observations, consisting of six 10.4-m antennas and nine 6.1-m antennas. CB\,26 was observed by CARMA on 2013 November 30 and 2013 December 02 with the antennas in B-configuration, providing a maximum baseline of 900 meters and a minimum baseline of $\sim$50 meters. At the central frequency of the observations, $\sim$102\,GHz (2.9\,mm), B-configuration yields an angular resolution of $\sim$0.7$''$ and a maximum angular scale of $\sim$10$''$. The correlator was configured in full continuum mode, with 8\,GHz of bandwidth subdivided into 16 windows with 500\,MHz of bandwidth. The quasar 0136+478 was used for bandpass calibration, 0359+509 was used as the complex gain calibrator, and Mars was used as the absolute flux calibrator. 

The data were calibrated and edited using the MIRIAD software package \citep{Sault1995} using standard methods. The reduced visibility data were then averaged to a single channel and imported into CASA for imaging and self-calibration. The data had sufficient signal to noise for self-calibration to be useful and we performed two rounds of phase-only self-calibration on the data, yielding only a 3\% improvement in maximum intensity and a 4\% improvement in noise level.


\subsection{VLA}

Observations of CB\,26 were conducted with the VLA in both A and B configuration (program 15A-381). In all observations, data were taken in both \Ka-band and C-band. The \Ka-band observations were conducted with the correlator in 3-bit mode, sampling a total of 8\,GHz of bandwidth that is separated into two 4\,GHz basebands. The two basebands were centered at 36.9\,GHz and 28.5\,GHz (8.1\,mm and 10.3\,mm). The C-band observations were conducted with the correlator in 8-bit mode, sampling a total of 2\,GHz of bandwidth. This mode was chosen because of the significant radio frequency interference (RFI) in C-band and the 8-bit samplers behaving better in the presence of RFI. Moreover, accounting for the lost bandwidth due to RFI, the 3-bit samplers would not yield significantly greater sensitivity than the 8-bit samplers. The two 1\,GHz basebands were centered at 4.7\,GHz and 7.3\,GHz (4.0\,cm and 6.4\,cm). For all observations, 3C84 was observed for bandpass calibration (\Ka-band only), 3C147 was observed for absolute flux calibration and C-band bandpass calibration. The complex gain calibrators for \Ka-band and C-band were J0458+5508 and J0541+5312, respectively. Different complex gain calibrators were necessary because J0458+5508 is resolved at C-band, but it was the nearest suitable calibrator for \Ka-band.

\subsubsection{B-configuration}

The B-configuration observations were executed on 2015 February 11 and were conducted in a single $\sim$4.5 hour scheduling block. B-configuration provides a maximum baseline of 11\,km, resulting in an effective resolution of $\sim$0.2$''$ at \Ka-band and 1$''$ in C-band. The \Ka-band observations were conducted in fast switching mode to compensate for short-timescale phase variations. The total time of a single on-source scan was 90\,sec, of which about 20\,sec was slew time. The time on the calibrator was 42\,sec, such that one complete cycle, calibrator, source, calibrator, took about 2.9\,min. The C-band observations did not require fast switching and each scan on CB\,26 was $\sim$600 seconds and $\sim$60 seconds on the calibrator. The total integration time on CB\,26 was $\sim$1.2\,hrs in \Ka-band and $\sim$1.15\,hrs in C-band.


\subsubsection{A-configuration}

Due to scheduling constraints, the A-configuration observations were conducted in three $\sim$1.6 hour scheduling blocks on 2015 September 04, September 23, and September 25. A-configuration provides a maximum baseline of 36\,km, resulting a resolution of $\sim$0.08$''$ at \Ka-band and $\sim$0.35$''$ in C-band. The last two executions were taken during the reconfiguration from A to D-configuration (most extended to most compact). However, the antennas on the longest baselines were moved last and there were still 20 antennas in A-configuration on September 23, and 17 antennas on September 25. We followed a similar observing strategy as B-configuration for the fast switching in \Ka-band and non-fast switching in C-band; however, we spent 480 seconds on CB\,26 in C-band rather than 600 seconds as in B-configuration. In each scheduling block, 0.41\,hrs was spent on CB\,26 in \Ka-band and 0.25hrs in C-band. Thus, the total time on source in \Ka-band was $\sim$1.25\,hrs and $\sim$0.75\,hrs in C-band.


\subsubsection{Data Reduction}

The raw visibility data were reduced using the scripted VLA pipeline version 1.3.1\footnote{\url{https://science.nrao.edu/facilities/vla/data-processing/pipeline/scripted-pipeline}} in CASA version 4.2.2 \citep{Mcmullin2007}. However, to obtain better accuracy in the flux calibration, we paused the pipeline at the flux calibration step to flag the flux calibration gain table, ensuring that no bad gain solutions were included in the flux calibration. Following the completed pipeline run, we flagged the gaintables for amplitude and phase and then used them as input to the \texttt{applycal} task using the option \texttt{mode=flagonly} to flag any science data that was associated with the flagged gain solutions. We also then inspected the calibrated science data in a time-averaged amplitude versus frequency plot to look for any obvious signs of bad data that was not excised with the gaintable flagging.

The \Ka-band images were produced using the CASA task \texttt{clean} using natural weighting for the combined A and B-configuration imaging. However, we included $uv$-distances $>$23\,k$\lambda$ in order to avoid including the antennas that had been moved to their D-configuration positions, which would negatively affect the synthesized beam. In the C-band images, we included $uv$-distances $>$10\,k$\lambda$.


\section{Model and modelling}
\label{sect_model_modelling}

To model the dust continuum data, we use the well-tested radiative transfer code \texttt{RADMC-3D}\footnote{\url{http://www.ita.uni-heidelberg.de/~dullemond/software/radmc-3d/}} \citep{Dullemond2012}. Our model is composed of a disk and an envelope, similar to the sketch in Fig.\,5 of \citet{Sauter2009}. Our employed disk model is the same as that in \citet{Sauter2009}, yet the envelope model is different from theirs. 


\subsection{Disk}
\label{sect_disk}

For the disk we assume a density profile as described by \citet{Shakura1973}:
\begin{equation}
\begin{aligned}
\rho_\disk(\vec{r}) & =  \rho_{\rm out,disk} \left( \frac{r_{\rm out}}{\rcyl} \right)^\alpha \exp \left(-\frac{1}{2}\left[ \frac{z}{h}\right]^2 \right) \\            
                    & = \frac{\Sigma(\vec{r})}{\sqrt{2\pi}h(\rcyl)} \exp \left(-\frac{1}{2}\left[ \frac{z}{h}\right]^2 \right), 
\label{equ_disk}
\end{aligned}
\end{equation}
where $h$ is vertical scale height as a function of the radial distance $\rcyl$ from the $z$--axis (see also eq.\,\ref{equ_scalehight}), $z$ is the cylindrical coordinate with $z=0$ corresponding to the disk midplane. The parameter $\rho_{\rm out,disk}$ is determined by the volume density at the outer radius $r_{\rm out}$, and $\Sigma(\vec{r})$ is the surface density at radius $\vec{r}$ \citep[see, e.g.,][]{Wolf2003,Stapelfeldt1998} as 
\begin{equation}
  \Sigma(\vec{r}) =\Sigma_{\rm out} \left( \frac{r_{\rm out}}{\rcyl} \right)^p.
  \label{equ_disk_sur}
\end{equation}
The function $h(\rcyl)$ is 
\begin{equation}
   h(\rcyl) = h_{\rm out} \left( \frac{\rcyl}{r_{\rm out}} \right)^k = \rcyl \frac{h_{\rm out}}{r_{\rm out}} \left( \frac{\rcyl}{r_{\rm out}} \right)^{k-1},
   \label{equ_scalehight}
\end{equation}
where $h_{\rm out}$ is the vertical scale height at the outer radius $r_{\rm out}$. 

The exponents $\alpha$ and $k$ in Equations (\ref{equ_disk}) and (\ref{equ_scalehight}) describe the radial density profile and disk flaring, respectively. Comparison with Equations (\ref{equ_disk}), (\ref{equ_disk_sur}) and (\ref{equ_scalehight}) yields the relation
\begin{equation}
  p = \alpha - k.
  \label{equ_p}
\end{equation}


Considering that the dust destruction radius is at $\sim$0.1\,au \citep{Eisner2005,Mann2010}, the modelling radii of the inner disk were set to \mbox{$0.1\lesssim\vec{r_{\rm in}}\lesssim100.0$\,au}, while the radii of the outer disk were set to \mbox{$100.0\lesssim\vec{r_{\rm in}}\lesssim300.0$\,au} \citep[cf.][]{Launhardt2009,Sauter2009,Akimkin2012}. This modelling strategy has already been successfully applied to various other edge-on disks, such as the Butterfly-Star IRAS\,04302+2247 \citep{Wolf2003}, HK\,Tau \citep{Stapelfeldt1998}, and HV\,Tau \citep{Wolf2003}.


\subsection{Envelope}
\label{sect_env}

For the model of the envelope, we use a spherically symmetric density distribution of the form:
\begin{equation}
   \rho_\env (\vec{r}) = \rho_{\rm out,\env} \left(\frac{\vec{r}}{R_{\rm out}}\right)^\gamma,
   \label{equ_env}
\end{equation}
where $\rho_{\rm out,\env}$ is determined by the dust density at the outer radius $R_{\rm out}$ of the envelope, and $\gamma$ is the radial density distribution power-law exponent.

To ensure a smooth and realistic density distribution, we do not simply add the two models, but substitute the envelope model (eq.\,\ref{equ_env}) with the disk model (eq.\,\ref{equ_disk}) at \mbox{$\rho_\disk(\vec{r}) \ge \rho_\env(\vec{r})$}. This implies that the dust density inside an inner hole in the disk is not zero, but is determined by the density profile of the envelope (cf. Fig.\,\ref{fig_density}). Thus, the final model of the density distribution could be described as
\begin{equation}
\rho(\vec{r}) = \left\{
   \begin{array}{r@{\quad:\quad}l}
   \rho_\disk(\vec{r}) & \rho_\disk(\vec{r}) \ge \rho_\env(\vec{r})  \\
   \rho_\env(\vec{r})  & \rho_\disk(\vec{r})  <  \rho_\env(\vec{r})
   \end{array} \right..
   \label{equ_dendiskenv}
\end{equation}
We also verified how adding the two density profiles instead of substituting them would affect our results and find that difference in the total model fluxes is less than 3\%, which is smaller than the uncertainties and thus not significant and negligible.
Considering that the dust destruction radius is at $\sim$0.1\,au \citep{Eisner2005,Mann2010}, the modelling radii of the envelope were set as $0.1\lesssim\vec{r}\lesssim900$\,au. We adopt 1100\,au as the radius of the modelled millimeter image space.





\subsection{Dust composition}
\label{sect_grain-size-distr}

\paragraph{Grain shape}
We assume the dust grains to be homogeneous spheres. We limit our model to be simple but also unambiguous by assuming spherical, non-aligned and non-orientated dust grains \citep{Voshchinnikov2002,Sauter2009}.

\paragraph{Grain composition}
For the chemical composition of the dust grains we employ a model that incorporates both silicate and graphite material \citep[see, e.g., ``Butterfly star'' in][]{Wolf2003}. For the optical data we use the complex refractive indices of smoothed astronomical silicate and graphite \citep{Weingartner2001}. Since these authors provide refractive indices only for wavelengths shorter than our modelled range, we extrapolate the refractive indices to the  wavelength of 6.4\,cm. For graphite we adopt the common ``$\frac{1}{3} - \frac{2}{3}$'' approximation. That means, if $Q_{\rm ext}$ is the extinction efficiency factor, then
\begin{equation}
   Q_{\rm ext, graph} = \frac{1}{3} Q_{\rm ext}(\epsilon_\parallel ) + \frac{2}{3} Q_{\rm ext}(\epsilon_\perp),.
   \label{equ_grainsize}
\end{equation}
where $\epsilon_\parallel$ and $\epsilon_\perp$ are the graphite dielectric tensor's components for the electric field parallel and orthogonal to the crystallographic axis, respectively. As has been shown by \citet{Draine1993}, this graphite model is sufficient for extinction curve modelling. For silicate, applying an abundance ratio from silicate to graphite of $3.9 \times 10^{-27} \cm^3 {\rm H}^{-1}$ : $2.3 \times 10^{-27} \cm^3 {\rm H}^{-1}$, we get relative abundances of $62.5 \%$ for astronomical silicate and $37.5\%$ graphite ($\frac{1}{3} \epsilon_\parallel$ and $\frac{2}{3}\epsilon_\perp$). The grain composition parameters are the same as those in \citet{Wolf2003} and \citep{Sauter2009}.

\paragraph{Grain sizes} 
The grain size distribution is assumed as a power law \citep{Mathis1977} in the form of
\begin{equation}
  n(a) \mathbf{\dd a} \sim a^{-3.5}  \mathbf{\dd a} \quad {\rm with} \quad  a_{\rm min} \le a \le a_{\rm max},
  \label{eqgraindist}
\end{equation}
where $a$ is the dust grain radius and $n(a)$ the number of dust grains with a specific radius. We assume an average grain mass density of \mbox{$\rho_{\rm grain} = 3.01 \,{\rm g}\,{\rm cm}^{-3}$}. We fix the minimum dust grain size as \mbox{$a_{\rm min}^{\rm disk}=a_{\rm min}^{\rm env}=10\nm$} \citep{Mathis1977} in the disk and envelope, respectively. The maximum grain size, $a_{\rm max}$, is set as a free parameter in the fitting and the choice of $a_{\rm max}$ in Fig.\,\ref{fig_dustkappa} is based on the best-fit model.


\subsection{Dust scattering}
\label{sect_scattering}

\begin{figure}[htp]
\centering
\includegraphics[width=0.45\textwidth, angle=0]{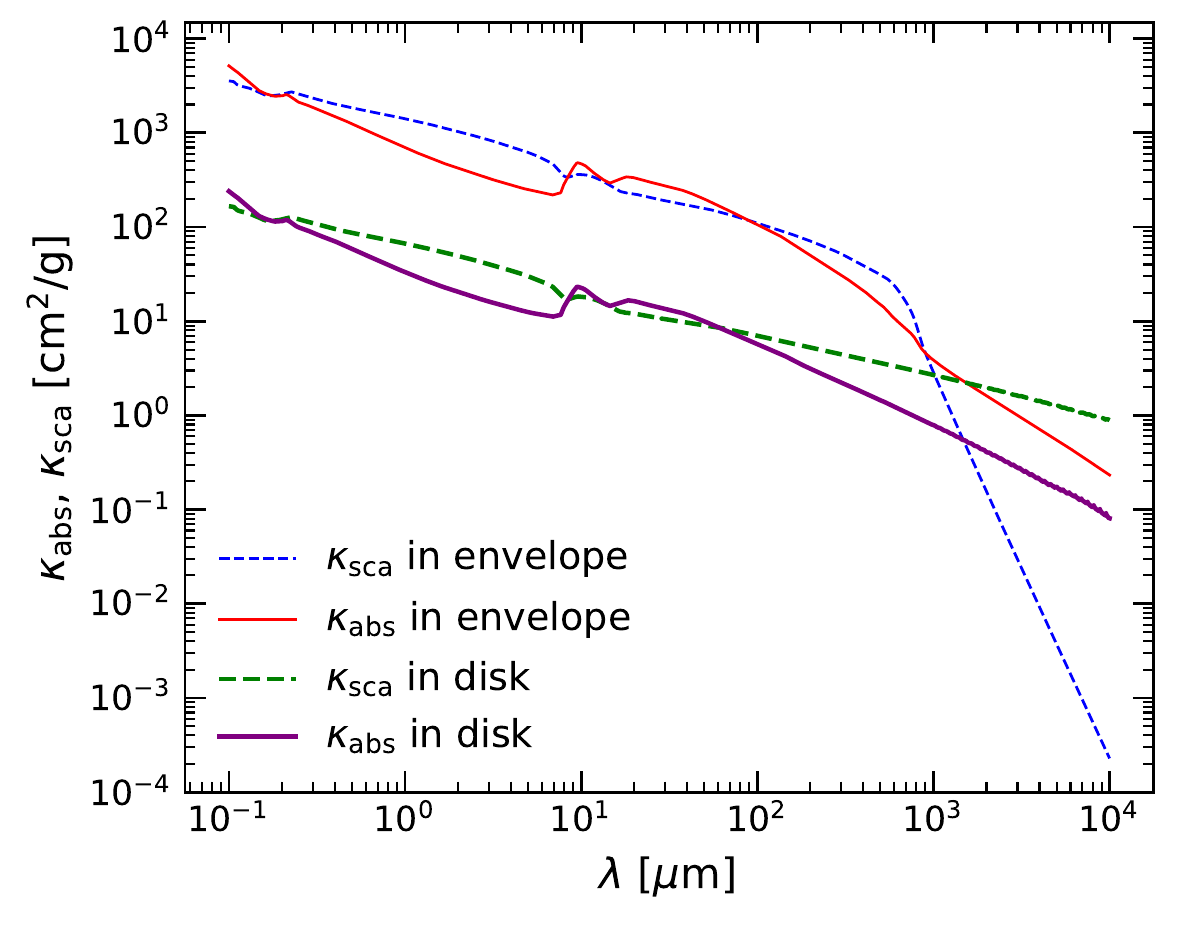}
\caption{Absorption and scattering coefficients as a function of wavelength for the dust grains in disk and envelope. The power-law slope $p$ of the grain size distribution is taken to be $p=-3.5$ with $a_{\rm max}^{\rm disk}=5\cm$, $a_{\rm max}^{\rm env}=110\mum$, and $a_{\rm min}^{\rm disk}=a_{\rm min}^{\rm env}=10\nm$.}
\label{fig_dustkappa}
\end{figure}

Light scattering by dust grains can considerably reduce the emission from an optically thick region and lead to an underestimation of the optical depth \citep{Zhu2019}. Scattering also affects the derivation of the maximum grain size of the dust distribution \citep[e.g.,][]{Kataoka2015,Carrasco2019}. Therefore, the effect of scattering in the dust opacity \citep{Natta1993} is included in our modelling of both the disk and envelope and for the entire wavelength range. Figure\,\ref{fig_dustkappa} shows the average absorption and scattering opacities of the disk and envelope over the whole grain size distribution. The power-law slope $p$ of the grain size distribution is taken to be $p=-3.5$, with maximum grain sizes of 
\mbox{$a_{\rm max}^{\rm disk}=5\cm$} and \mbox{$a_{\rm max}^{\rm env}=110\mum$}, and a  minimum grain size of 
\mbox{$a_{\rm min}^{\rm disk}=a_{\rm min}^{\rm env}=10\nm$}. 

If scattering is ignored, we can no longer fit the infrared data points at $\lambda\lesssim20\mum$ (cf. Fig.\,\ref{fig_sed}). The mm SED of CB\,26 can still be reproduced without scattering, albeit with a slightly different result on the best-fitting value of $a_{\rm max}$ (see Sect.\,\ref{sect_fit} and discussion in Sect.\,\ref{sect_growth}).


\subsection{Stellar heating}

Due to the edge-on orientation of the disk, the central star cannot be observed directly. We thus have to assume the stellar parameters. Observations only hint at a luminosity being $L_\star \ge 0.5 \,\lsun$ \citep{Stecklum2004}. As a starting point we choose a typical T\,Tauri star as described by \citet{Gullbring1998}. This star has a radius of $r_\star=2$\,\mbox{\rsol} and a luminosity of $L_\star = 0.92 \,L_\odot$ \citep[cf.][]{Launhardt2009}. Assuming the star to be a black body radiator, this yields an effective surface temperature of $T_\star = 4000\,\K$. We thus set the temperature as a varied parameter around 4000\,K in the models. The star will be the only primary source of energy, which means that the disk is passive. The stellar radiation heats the dust, which in turn itself re-emits at longer wavelengths. We neglect accretion or turbulent processes within the models as other possible energy sources.


\subsection{Extinction}
\label{sect_extinction}

Extinction has a serious impact on the SED, especially at ultraviolet, optical, and infrared wavelengths \citep[e.g.,][]{Cardelli1989,Indebetouw2005}. 
The interstellar extinction in the wavelength range of \mbox{$0.85 \la \lambda \la 11.6\mum$} was assumed to follow the relationship described by \citet{Cardelli1989}. Therefore, we can use $A_{\rm V}$ as a free parameter in the SED fitting that accounts for both the envelope of CB\,26 and interstellar dusty material along the line of sight.


\subsection{Fitting}
\label{sect_fit}

Since the protostellar disk in CB\,26 is close to edge-on with an inclination angle of $\sim$88$^\circ$ \citep{Sauter2009,Launhardt2009,Launhardt2010}, the emission is extremely non-spherically symmetric \citep{Jorgensen2009}. Fitting averaged 1D visibility curves could be misleading in such a case. Unfortunately, the extreme edge-on configuration also makes deprojection impossible. Therefore, we only consider the 2D visibilities rather than 1D visibility curves, although the signal-to-noise ratio per visibility point is much lower. We therefore compare the simulations with the observations only in image plane instead of the ($u, v$)-plane\footnote{We have examined the 2D visibilities of 1.3\,mm in the best-fit model and found them to agree well with the observations.}. We follow three steps to fit the data as outlined below:

\begin{enumerate}
\item We first fit the SED alone, by a manual exploration of the parameter space, to derive a preliminary estimate of the main parameters. The goal is to determine which parameters can be kept fixed (see Table\,\ref{tab_photometric}), and which parameters need to be varied in our modelling, thus allowing us to significantly reduce the dimensionality of the parameter space to be explored \citep{Pinte2008};

\item We then systematically explore these remaining free parameters, which cannot be easily extracted from the observations and/or can be correlated with each other. We calculate a rough grid of models and perform a simultaneous fit to all observations. The aim is to find the models which could roughly fit the observations and to narrow down the parameter space;

\item We finally calculate a fine grid of models in the narrowed-down parameter space (in step\,2) to fit the observations. This also reduces the uncertainties of the modelled parameters.
\end{enumerate}

In total, more than 300,000 models are performed for fitting.
For each comparison between model and observation, we obtain four individual  $\chi^2_{\rm i}$s; one value for the SED and three values for the millimeter maps at 1.3, 2.9, and 8.1\,mm. To measure the fluxes in our model maps for comparison with the SED data, we apply circular apertures with the sizes listed Table\,\ref{tab_photometric}.) 
We do not consider the 1.1\,mm SMA map in the modelling because of the large flux calibration uncertainties (see Sects.\,\ref{sect_obs_sma} and \ref{sect_chi2}). We also exclude the maps at $\lambda>8.1$\,mm from the modelling because they are dominated by free-free emission and have low SNR. To fit 8.1\,mm map, we mask the central bright area ($\gtrsim3\sigma$ in the 8.1\,mm residual map of Fig.\,\ref{fig_obs_sim_2d}) because its flux is strongly contaminated by free-free emission (see detailed discussion in Sect.\,\ref{sect_free-free}). To calculate individual $\chi^2_{\rm i}$s, we use the uncertainties listed in Table\,\ref{tab_photometric} and in the caption of Fig.\,\ref{fig_obs_sim_2d}, at each wavelength for the SED and the millimeter maps, respectively. 
To avoid fitting into the noise, we define for each of the mm maps a  polygon that includes the outermost closed $3\sigma$ contour, and consider only the flux measurements from inside these polygons.
The total ${\chi}^2_{\rm total}$ of one model is then the average over all the individual ${\chi}^2_{\rm i}$s:
\begin{equation}
   {\chi}^2_{\rm total} = \frac{1}{4}\left({\chi}^2_{\rm SED} + {\chi}^2_{\rm1.3mm} + {\chi}^2_{\rm2.9mm} + {\chi}^2_{\rm8.1mm}\right).
  \label{eq_sumxi}
\end{equation}
Based on the derived ${\chi}^2_{\rm total}$ and the best-fit model ${\chi}^2_{\rm best}$, the modelling uncertainties (minimum and maximum) are estimated by altering each free parameter value alone for the set of models that fulfill the ${\chi}^2_{\rm total}-{\chi}^2_{\rm best}<3$ criterion \citep[see a similar criterion in][]{Robitaille2007}. Although the choice of this level is somewhat arbitrary, it provides a range of acceptable fits to the eye, and reflects a dependence on each free parameter. All the input modelling parameter space and the output best-fit parameters are listed in Table\,\ref{tab_model}.

\begin{table}[htp]
\caption{Overview of parameter ranges used in the modelling.}
\label{tab_model} 
\centering \small  
\setlength{\tabcolsep}{1.6mm}{
\begin{tabular}{lc|ccc}
\hline \hline
Parameter &  Unit & Range & Best fit & Uncertainties  \\
\hline
\multicolumn{2}{c}{Disk} \\
$r_{\rm in}$               & au     & $0.1\sim100.0$            & 16.0     & $^{+37.4}_{-8.0}$       \\
$r_{\rm out}$              & au     & $100.0\sim300.0$          & 172.0    & $^{+20.4}_{-23.1}$       \\
$h_{\rm out}/r_{\rm out}$  &        & $0.05\sim0.50$            & 0.22     & $^{+0.06}_{-0.06}$      \\
$p$                    &        & $0.5\sim2.0$                 & $0.87$   & $^{+0.38}_{-0.32}$      \\
$k$           &        & $1.0\sim5.0$                       & 1.185    & $^{+0.190}_{-0.160}$      \\
$\Sigma_{\rm out}$  & $\rm g\,cm^{-2}$  & $(0.1\sim10)\times10^{-2}$ & $4.6\times10^{-2}$ & $^{+1.3}_{-1.2}\times10^{-2}$ \\
$\phi$                     & deg    & $60.0\sim90.0$            & 88.0     & $^{+2.0}_{-5.2}$       \\
P.A.                       & deg    & $-20.0\sim-40.0$          & $-32.0$  & $^{-5.5}_{+5.8}$       \\
\multicolumn{2}{c}{Envelope} \\
$R_{\rm in}$               & au     & 0.1                       & 0.1      & Fixed     \\
$R_{\rm out}$              & au     & 900                       & 900      & Fixed     \\
$\rho_{\rm out}$    & $\rm g\,cm^{-3}$     & $(0.1\sim90)\times10^{-20}$  & $6.1\times10^{-20}$ & $^{+3.7}_{-2.4}\times10^{-20}$ \\
$\gamma$                   &        & $-0.01\sim-0.5$           & $-0.10$   & $^{-0.17}_{+0.10}$     \\
P.C.                       & deg    & 40.0                      & 40.0     & Fixed     \\
\multicolumn{2}{c}{Star} \\

$r_{\star}$                & \rsol  & 2.0                    & 2.0      & Fixed     \\
$M_{\star}$                & \msol   & 0.55                   & 0.55     & Fixed     \\
$T_{\star}$                & K      & $3000\sim5000$            & 4000     & $^{+248}_{-340}$        \\
$A_{\rm V}$                & mag    & $1.0\sim30.0$             & 13.06    & $^{+2.40}_{-1.61}$       \\
Distance                   & pc     & 140.0                     & 140.0    & Fixed     \\
\multicolumn{2}{c}{Grain size distribution} \\
$a_{\rm max}^{\rm disk}$   & $\mu$m & $0.25\sim100\,000$        & 50\,000  & $^{+39\,000}_{-20\,400}$    \\
$a_{\rm max}^{\rm env}$    & $\mu$m & $0.25\sim100\,00$         & 110      & $^{+219}_{-69}$     \\
$a_{\rm min}^{\rm disk}$   & $\mu$m & 0.01                     & 0.01    & Fixed     \\
$a_{\rm min}^{\rm env}$    & $\mu$m & 0.01                     & 0.01    & Fixed     \\
\hline
\end{tabular}}
\begin{flushleft}
\textbf{Notes.} \\
$r_{\rm in}$: Inner radius of the disk.   \\
$r_{\rm out}$: Outer radius of the disk.   \\
$h_{\rm out}/r_{\rm out}$: Ratio of the pressure scale height to the radius at $r_{\rm out}$.   \\
$p$: Power exponent of the radial surface density distribution.   \\
$k$: Flare index.   \\
$\Sigma_{\rm out}$: Surface density of the disk at the outer radius.   \\
$\phi$: Inclination angle of the disk.   \\
$R_{\rm in}$: Inner radius of the envelope.   \\
$R_{\rm out}$: Outer radius of the envelope.   \\
$\rho_{\rm out}$: Density of the envelope at the outer radius.   \\
$\gamma$: Radial density distribution power exponent.   \\
P.C.: Radius of the polar cavity in the envelope.   \\
$r_{\star}$: Radius of the star.   \\
$M_{\star}$: Mass of the star, fitted with a Keplerian disk in the $^{12}$CO $J=2-1$ velocity field \citep{Launhardt2020}.  \\
$T_{\star}$: Effective surface temperature of the star.   \\
P.A.: The rotation of the major axis in the image plane.   \\
$A_{\rm V}$: Visual extinction. \\
Distance: The distance to the Sun.   \\
$a_{\rm max}^{\rm disk}$: Maximum dust grain size in the disk.   \\
$a_{\rm max}^{\rm env}$: Maximum dust grain size in the envelope.   \\
$a_{\rm min}^{\rm disk}$: Minimum dust grain size in the disk.   \\
$a_{\rm min}^{\rm env}$: Minimum dust grain size in the envelope. \\
Uncertainties: Defined by ${\chi}^2_{\rm total}-{\chi}^2_{\rm best}<3$.
\end{flushleft}
\end{table}


\section{Results}
\label{sect_results}


\subsection{Comparison between observation and simulation}
\label{sect_chi2}

\begin{figure*}
\centering
\includegraphics[width=0.85\textwidth, angle=0]{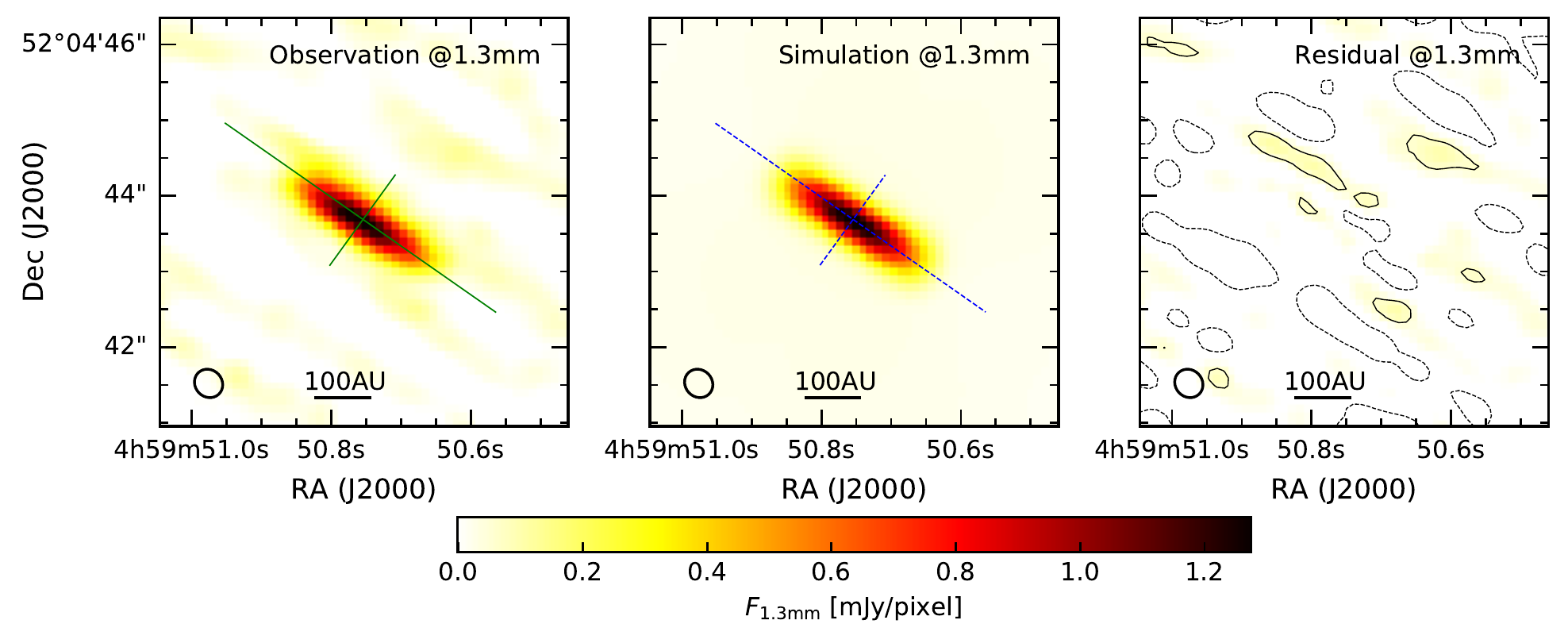}
\includegraphics[width=0.85\textwidth, angle=0]{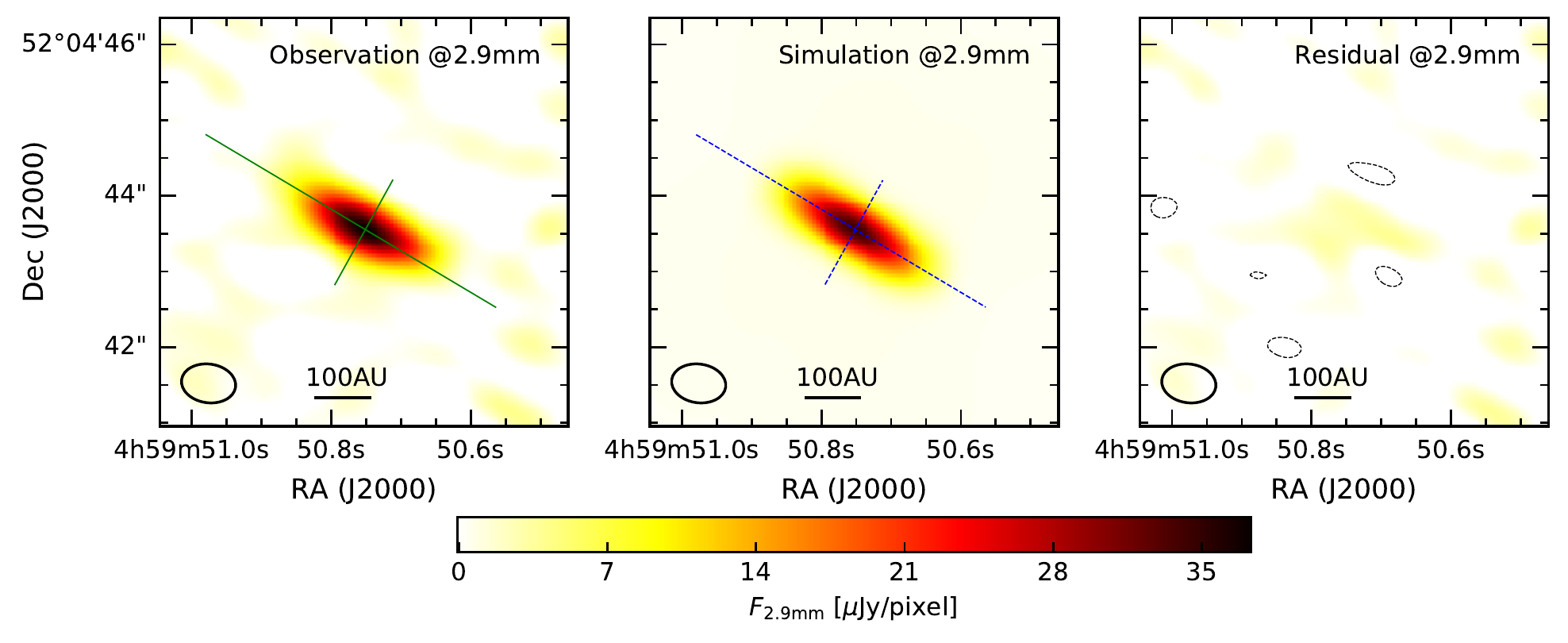}
\includegraphics[width=0.85\textwidth, angle=0]{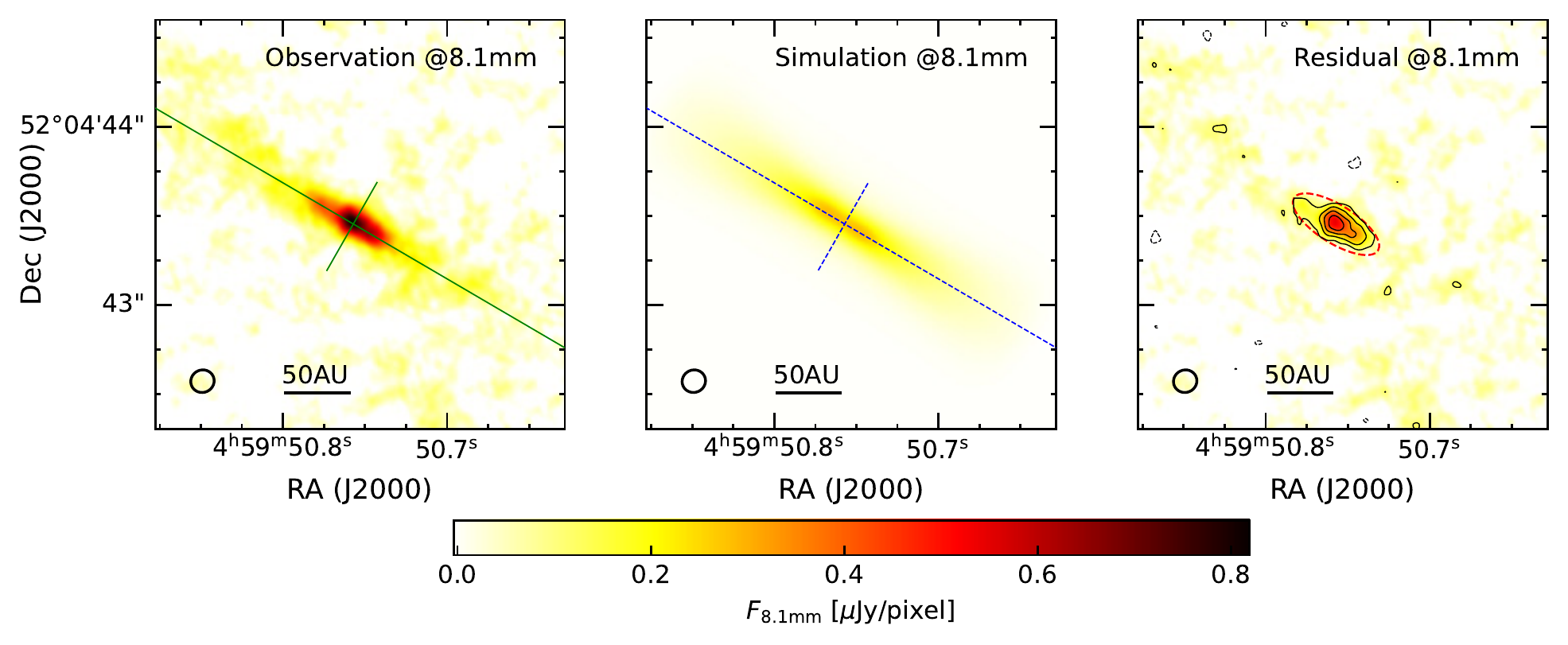}
\caption{Observed and modelled intensity maps at 1.3, 2.9, and 8.1\,mm. The modelled maps have smaller pixel sizes than the observed maps, but are convolved to the same beam sizes as the respective observations (shown as ellipses in the lower left corners). The contour levels in each residual image start at -3$\sigma$, 3$\sigma$ in steps of 3$\sigma$ with $\sigma_{\rm 1.3mm}=0.019\,\mjyp$ ($\rm 1\,pixel=0.1''\times0.1''$ at 1.3\,mm), $\sigma_{\rm 2.9mm}=1.37\,\mujyp$ ($\rm 1\,pixel=0.05''\times0.05''$ at 2.9\,mm), and $\sigma_{\rm 8.1mm}=0.038\,\mujyp$ ($\rm 1\,pixel=0.01''\times0.01''$ at 8.1\,mm) in the observations. Solid green and dotted blue lines show the directions of major and minor axes in the observation and simulation panels (cf. Fig.\,\ref{fig_obs_sim_1d_app}). The central compact emission within the red-dashed ellipse at 8.1\,mm residual map was masked for fitting (see details in Sect.\,\ref{sect_fit}).}
\label{fig_obs_sim_2d}
\end{figure*}

\begin{figure*}
\centering
\includegraphics[width=0.80\textwidth, angle=0]{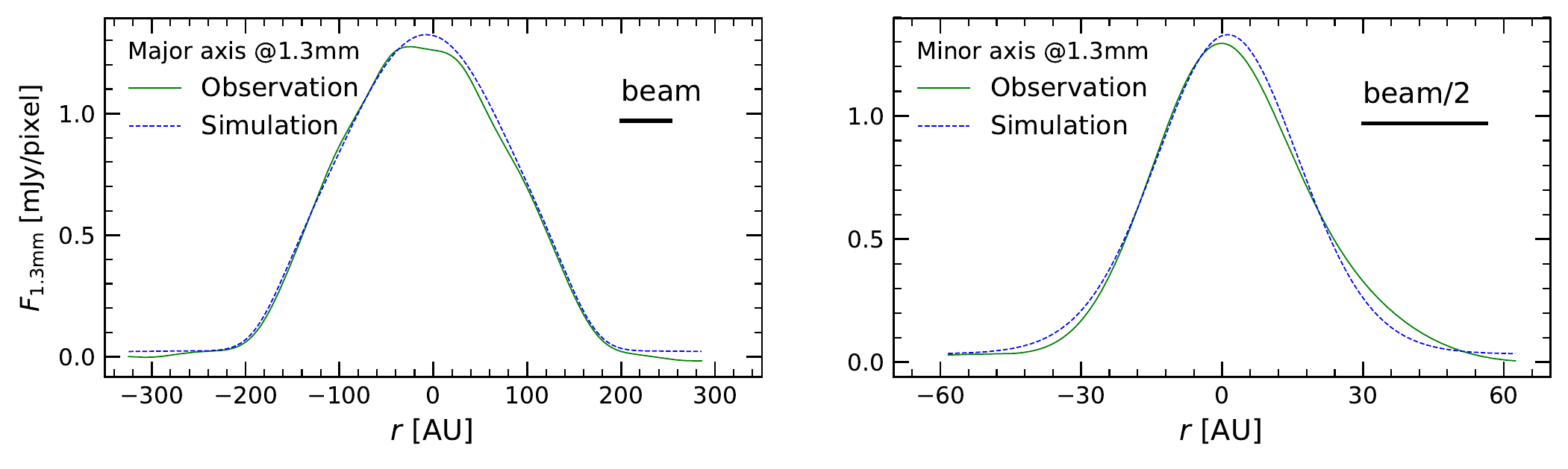}
\includegraphics[width=0.80\textwidth, angle=0]{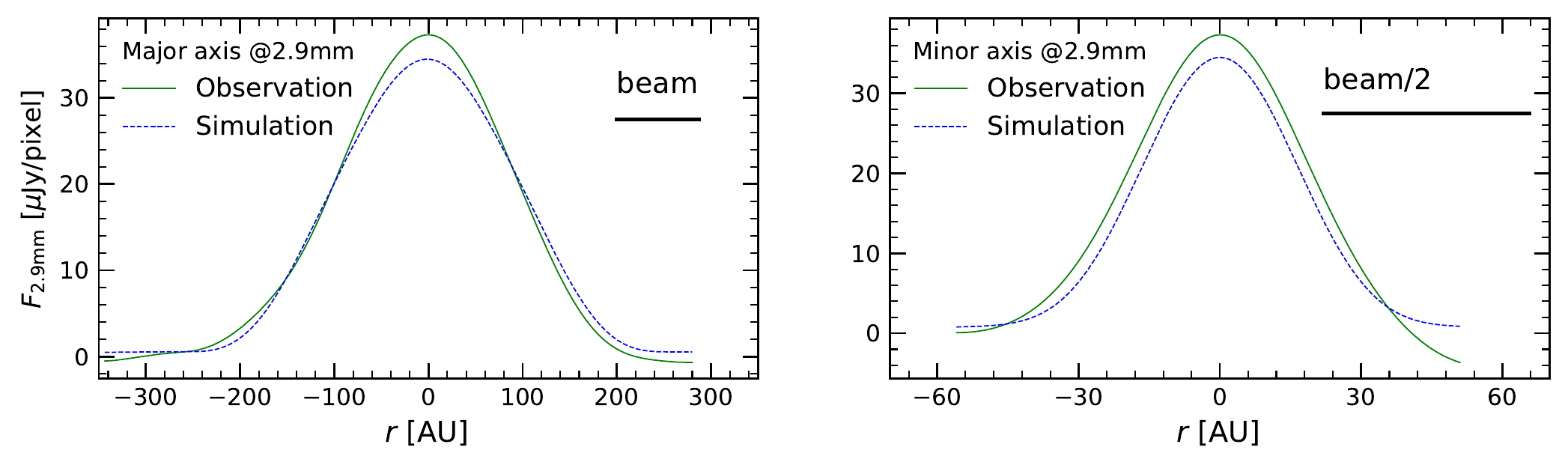}
\includegraphics[width=0.80\textwidth, angle=0]{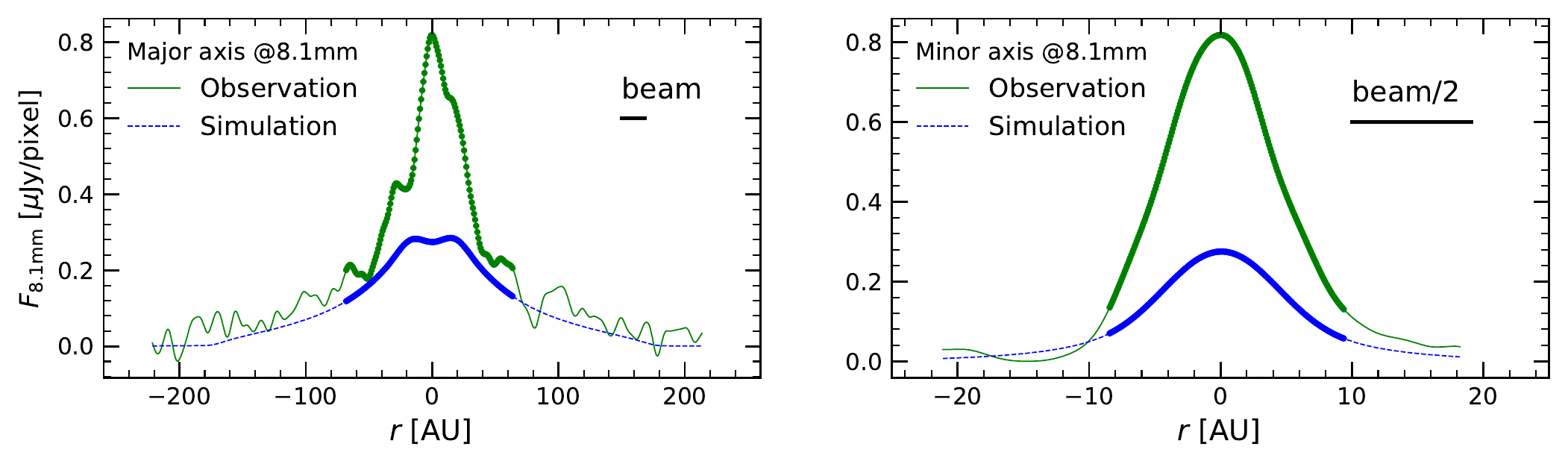}
\includegraphics[width=0.80\textwidth, angle=0]{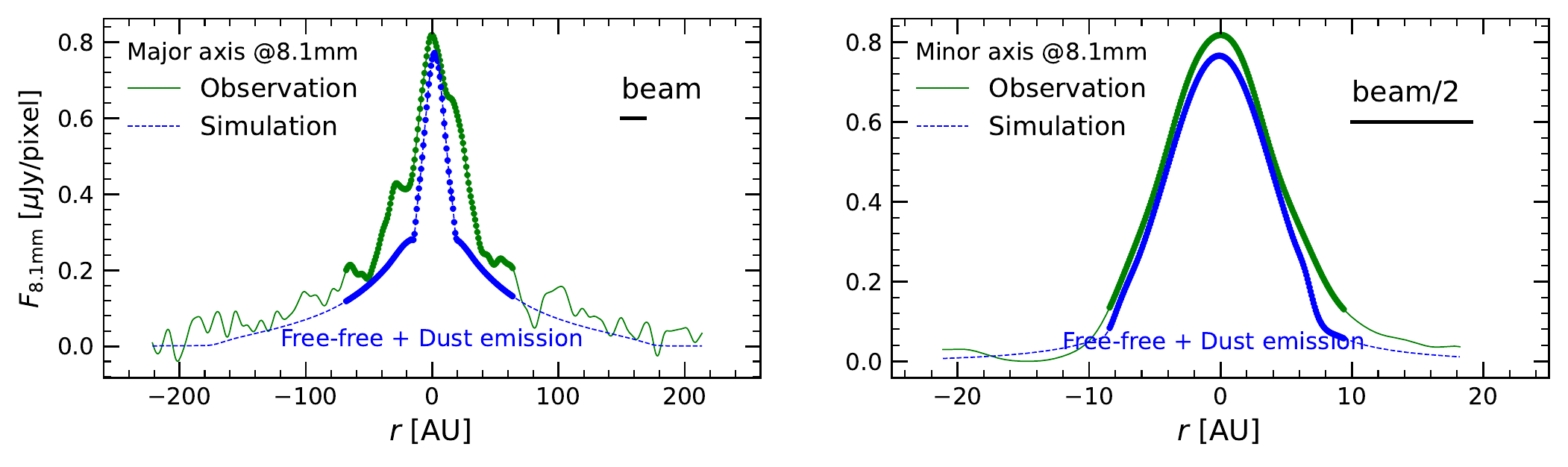}
\caption{Intensity distribution cuts along the major and minor axes (see Fig.\,\ref{fig_obs_sim_2d}) in observations and simulations at 1.3, 2.9, and 8.1\,mm, respectively. The pixel size at each wavelength is listed in the caption of Fig.\,\ref{fig_obs_sim_2d}. The central peak at 8.1\,mm was masked for fitting (marked with thick line, see details in Sect.\,\ref{sect_fit} and Fig.\,\ref{fig_obs_sim_2d}). In the lowest panel, 8.1\,mm free-free emission is included by assuming a point source with flux $\sim$77\,$\mu$Jy (see Sect.\,\ref{sect_free-free}). }
\label{fig_obs_sim_1d}
\end{figure*}

\begin{table}[htp]
\caption{${\chi}^2$ distributions for the best-fit model.}
\label{tab_chi2} \centering \small  
\setlength{\tabcolsep}{1.6mm}{
\begin{tabular}{l|cc}
\hline \hline
                         &  Without 1.1\,mm  &  With 1.1\,mm  \\
\hline
${\chi}^2_{\rm SED}$     &  2.45              & 2.38   \\
${\chi}^2_{\rm 1.1mm}$   &  ---               & 12.73  \\
${\chi}^2_{\rm 1.3mm}$   &  5.22              & 7.81   \\
${\chi}^2_{\rm 2.9mm}$   &  2.07              & 1.63   \\
${\chi}^2_{\rm 8.1mm}$   &  1.40              & 1.33   \\
${\chi}^2_{\rm total}$   &  2.79              & 5.18   \\
\hline
\end{tabular}}
\end{table}

The total ${\chi}^2_{\rm total}$ for the best-fit model to the SED and the three millimeter maps at 1.3, 2.9, and 8.1\,mm is \mbox{${\chi}^2_{\rm total} = 2.79$} (Table\,\ref{tab_chi2}). Figure\,\ref{fig_sed} shows the SED between 0.9$\mum$ and 6.4\,cm together with this best fit model SED, which fits the observations quite well with ${\chi}^2_{\rm SED} = 2.45$. The sudden flux decrease at $\lambda\sim$1\,mm in the SED (Fig.\,\ref{fig_sed}) is related to the inability of the interferometric observations at longer wavelengths to recover the total flux from the extended envelope. The fluxes at these wavelengths are therefore measured in smaller apertures (Table\,\ref{tab_photometric}) that contain only the disk emission, as opposed to the total disk+envelope flux measured at the shorter wavelengths. 

Figure\,\ref{fig_obs_sim_2d} shows the observed and modelled intensity maps at 1.3, 2.9, and 8.1\,mm. The maps of the best-fit model at 1.3, 2.9, and 8.1\,mm have \mbox{${\chi}^2_{\rm 1.3mm}=5.22$}, \mbox{${\chi}^2_{\rm 2.9mm}=2.07$}, and \mbox{${\chi}^2_{\rm 8.1mm}=1.40$}, respectively. The residual map at 1.3\,mm shows some non-axisymmetric $\pm3\sigma$ contours, which are probably phase noise or deconvolution artifacts. Figure\,\ref{fig_obs_sim_1d} displays the intensity distribution cuts along the major and minor axes in the observed and simulated maps at 1.3, 2.9, and 8.1\,mm, respectively (see Fig.\,\ref{fig_obs_sim_2d}). The discrepancies at all three wavelengths (at 8.1\,mm considering only regions outside the masked central free-free peak) are smaller than $3\sigma$. 

If the 1.1\,mm is included for calculating the best-fit model, the total ${\chi}^2_{\rm total}$ will be ${\chi}^2_{\rm total,\,with\,1.1mm} = 5.18$\footnote{${\chi}^2_{\rm total,\,with\,1.1mm} = \frac{1}{5}\left({\chi}^2_{\rm SED} + {\chi}^2_{\rm1.1mm}  + {\chi}^2_{\rm1.3mm} + {\chi}^2_{\rm2.9mm} + {\chi}^2_{\rm8.1mm}\right)$}, which is almost twice as high as when excluding it. The individual ${\chi}^2$ values listed in Table\,\ref{tab_chi2} indicate that the fits to both the 1.1\,mm and the 1.3\,mm are particularly bad, with \mbox{${\chi}^2_{\rm1.1mm}=12.73$} and \mbox{${\chi}^2_{\rm1.3mm}=7.81$}, when the 1.1\,mm map is included. This means, these two maps cannot be fit well together, i.e., they disagree. We suspect the main reason for this discrepancy to be the poor flux calibration of the 1.1\,mm SMA data (see Sects.\,\ref{sect_obs_sma} and \ref{sect_fit}). The observed and modelled intensity maps for the fit including the 1.1\,mm map are shown in appendix Fig.\,\ref{fig_obs_sim_2d_app} (corresponding to Fig.\,\ref{fig_obs_sim_2d}), and their intensity distribution cuts along the major and minor axes are presented in appendix Fig.\,\ref{fig_obs_sim_1d_app} (corresponding to Fig.\,\ref{fig_obs_sim_1d}).
Nevertheless, the best-fit parameter values listed in Table\,\ref{tab_model_app} all agree within the uncertainties with the results obtained without the 1.1\,mm map and the constraints on the maximum grain size in the disk are completely unaltered. Therefore, we adopt the best-fit model obtained without the 1.1\,mm map for further discussion\footnote{The SED slope of the data points between 1.3 and 2.9\,mm is \mbox{$\alpha_{\rm mm} = -2.63\pm0.40$} (see Fig.\,\ref{fig_sed} and Table\,\ref{tab_photometric}). This indicates that the 1.1\,mm data point ($F_{\rm 1.1mm}=238\pm45$) lies within $\sim$1.2$\sigma$ on that slope, despite the aferomentioned increased uncertainty for the 1.1\,mm data.}.



\subsection{Density}
\label{sect_density}

\begin{figure}[htp]
\centering
\includegraphics[height=0.36\textwidth, angle=0]{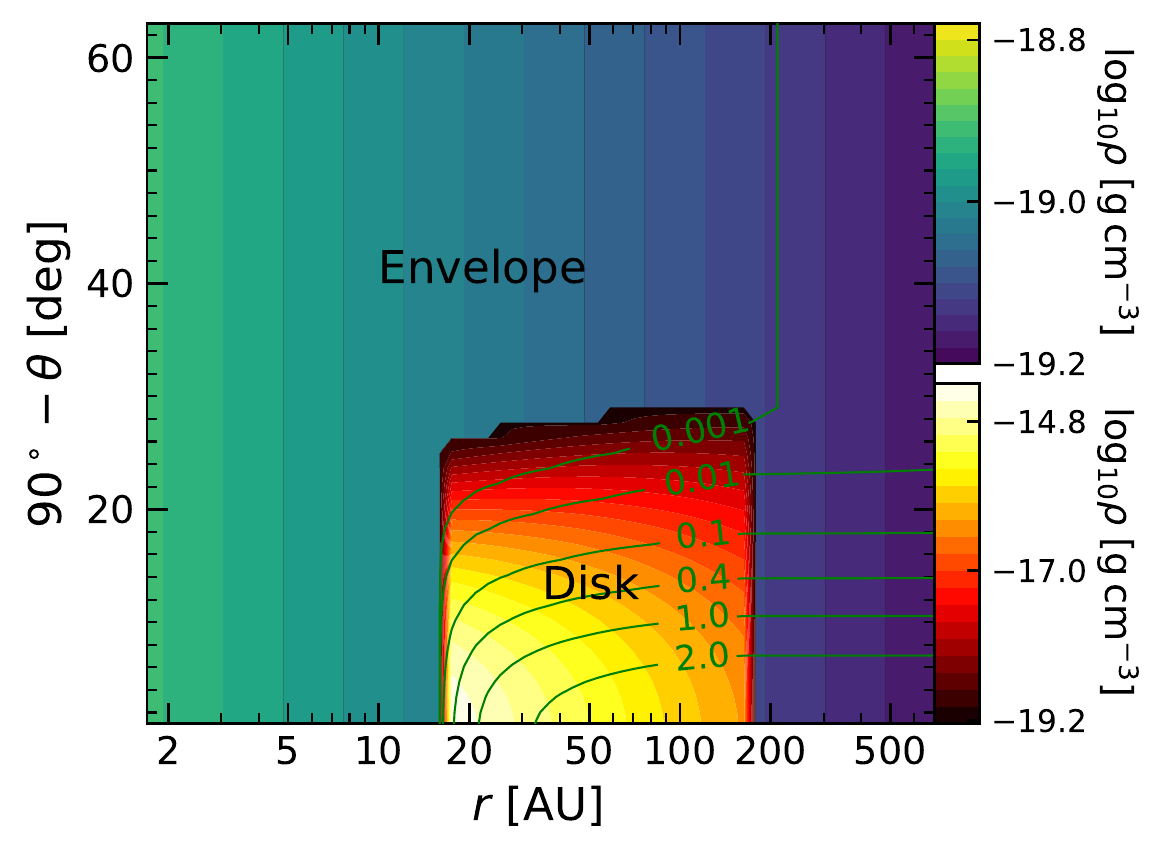}
\caption{The density structure as a function of radius $r$ and polar angle $\theta$ in spherical coordinates (90$^\circ$ is the equatorial plane with $z=0$). The disk and envelope locations are indicated by different color scales. The contours with numbers indicate different optical depths at 1.3\,mm.}
\label{fig_density}
\end{figure}

Figure\,\ref{fig_density} displays the density structure as a function of radius $r$ and polar angle $\theta$ in spherical coordinates (90$^\circ$ is the equatorial plane with $z=0$). It clearly shows the density structures in the disk (see Equation\,(\ref{equ_disk})) and envelope (see Equation\,(\ref{equ_env})). The disk has an inner hole like already found before \citep{Launhardt2009,Sauter2009,Akimkin2012}. Although our best-fit hole radius of $r_{\rm in} = 16.0^{+37.4}_{-8.0}$\,au is somewhat smaller than derived before, all four hole size estimates agree within the uncertainties. The maximum dust density in the disk is $\rho_{\rm max} \approx 3.5\times10^{-15}\,{\rm g\,cm^{-3}}$, located at the inner edge of the disk with $r_{\rm in} = 16.0$\,au. We also overlay the distribution of optical depths at 1.3\,mm on the density map. The regions close to midplane with $90^\circ-\theta \la 10^\circ$ have moderate optical depth with $\tau_{\rm 1.3mm}\ga1.0$.


\subsection{Temperature}
\label{sect_temperature}

\begin{figure}[htp]
\centering
\includegraphics[height=0.36\textwidth, angle=0]{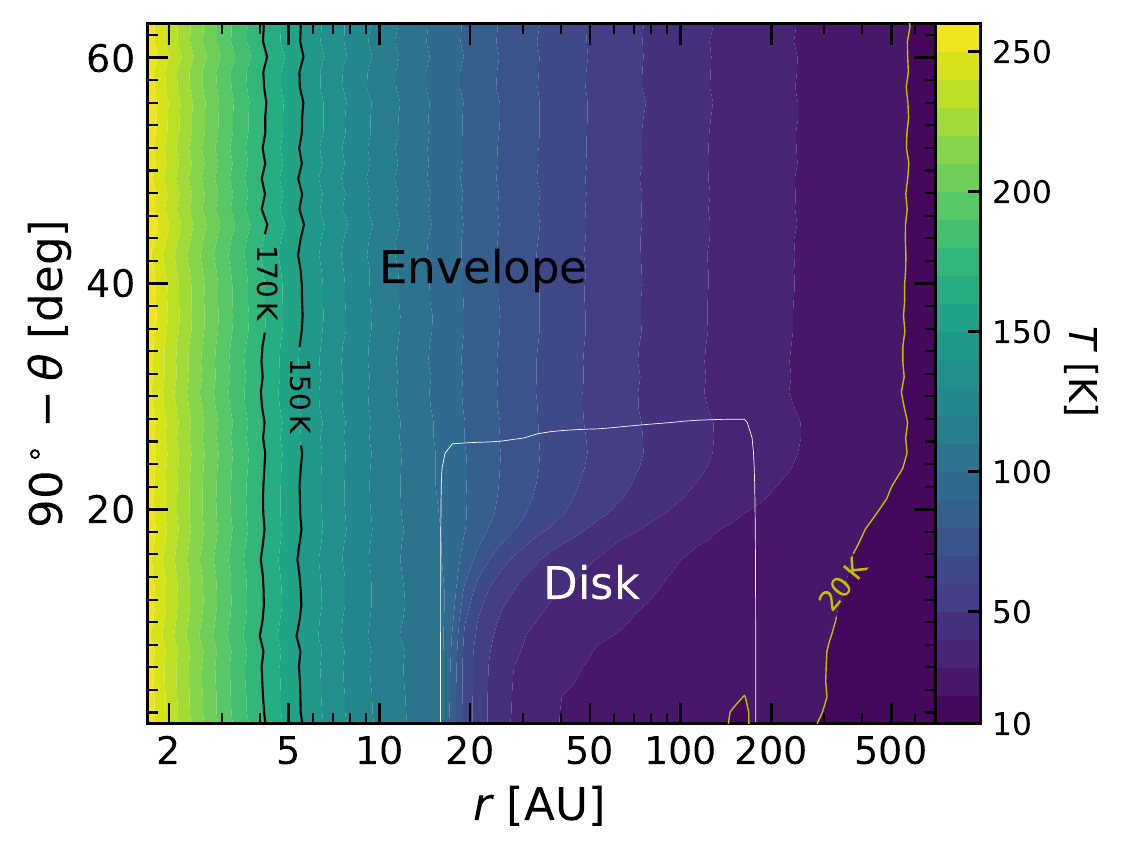}
\caption{The temperature structure as a function of radius $r$ and polar angle $\theta$ in spherical coordinates. The possible CO snow line at $\sim$20\,K and the water ice sublimation temperature range of 150\,--\,170\,K are indicated with three horizontal lines, respectively.}
\label{fig_temperature}
\end{figure}

\begin{figure}[htp]
\centering
\includegraphics[width=0.45\textwidth, angle=0]{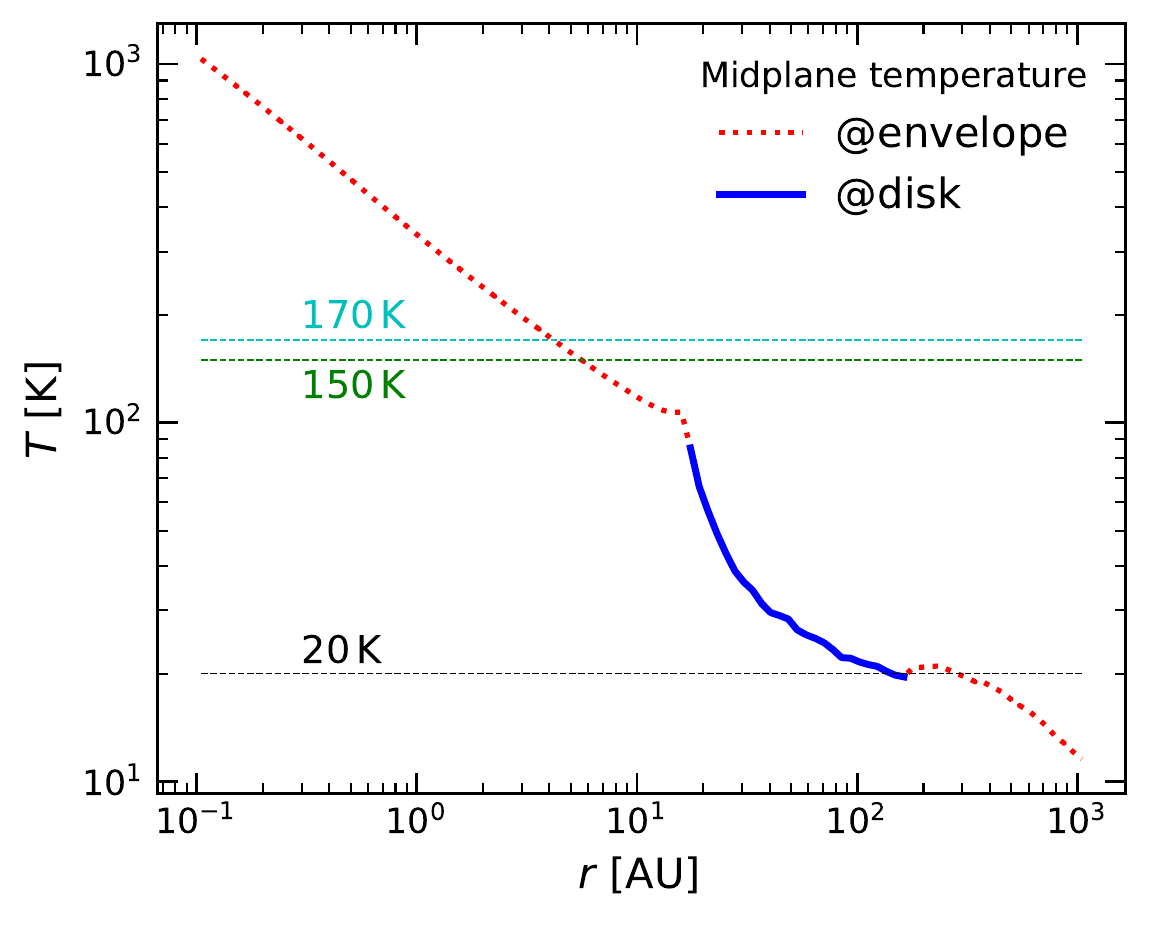}
\caption{Temperature distribution along the disk and envelope midplane. The possible CO snow line of 20\,K and ice sublimation temperature range of 150\,--\,170\,K are indicated with three dashed lines, respectively.}
\label{fig_temperature_1d}
\end{figure}

\begin{figure*}[htp]
\centering
\includegraphics[width=0.33\textwidth, angle=0]{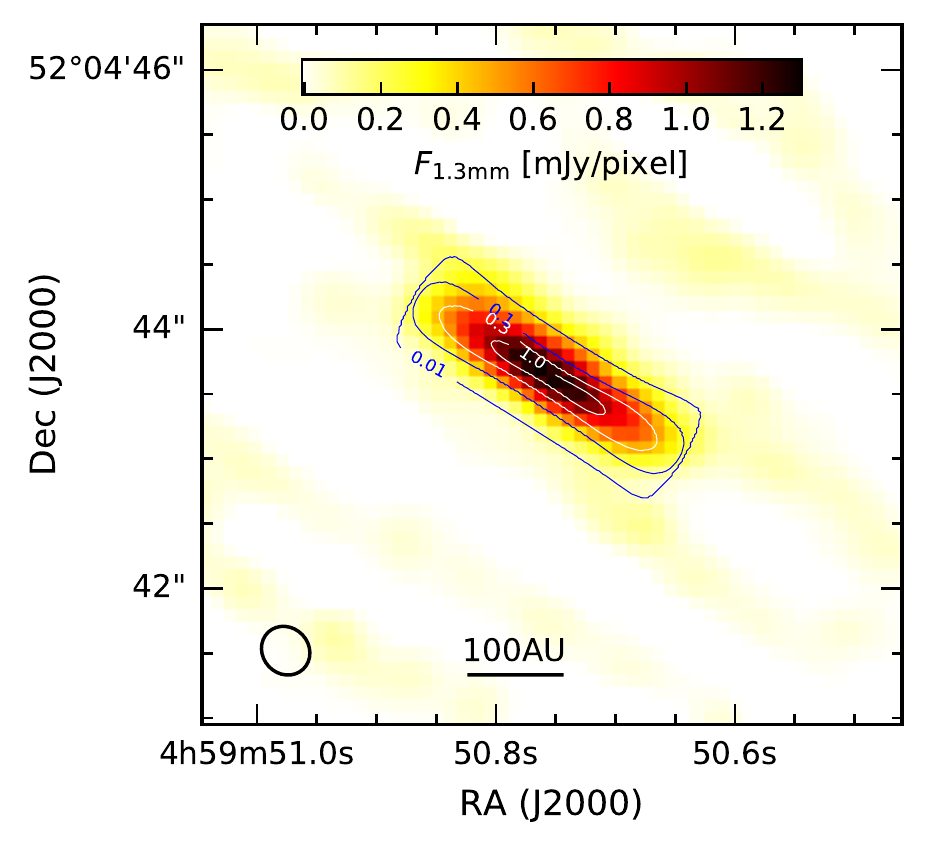} 
\includegraphics[width=0.33\textwidth, angle=0]{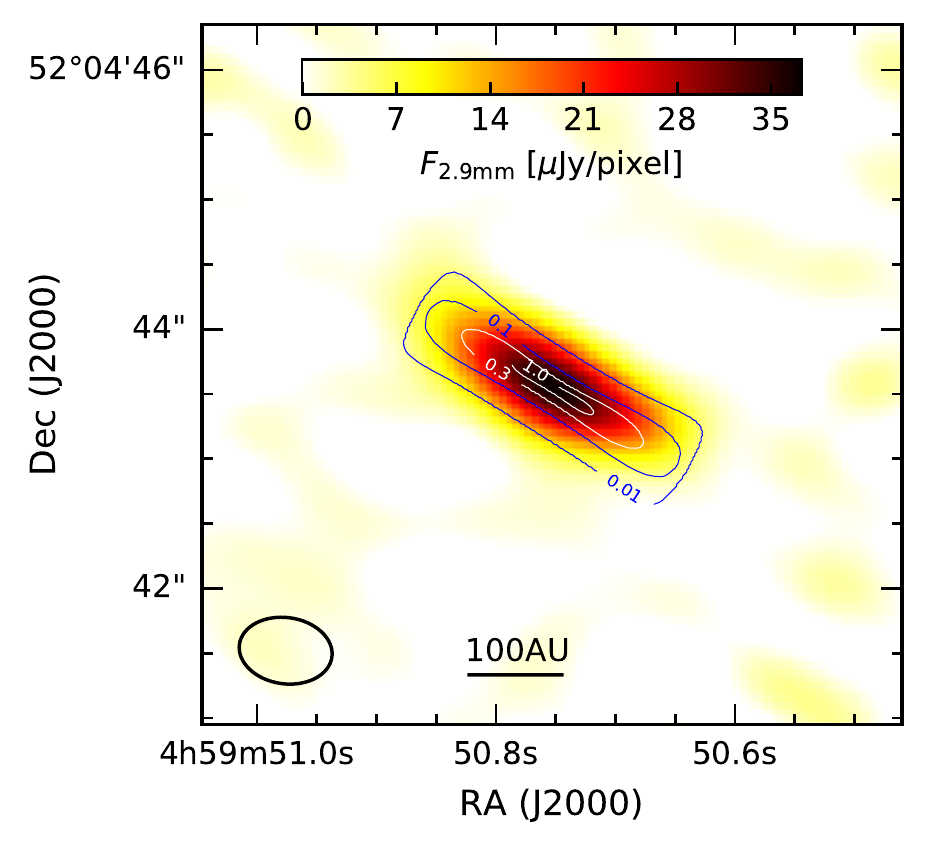}
\includegraphics[width=0.33\textwidth, angle=0]{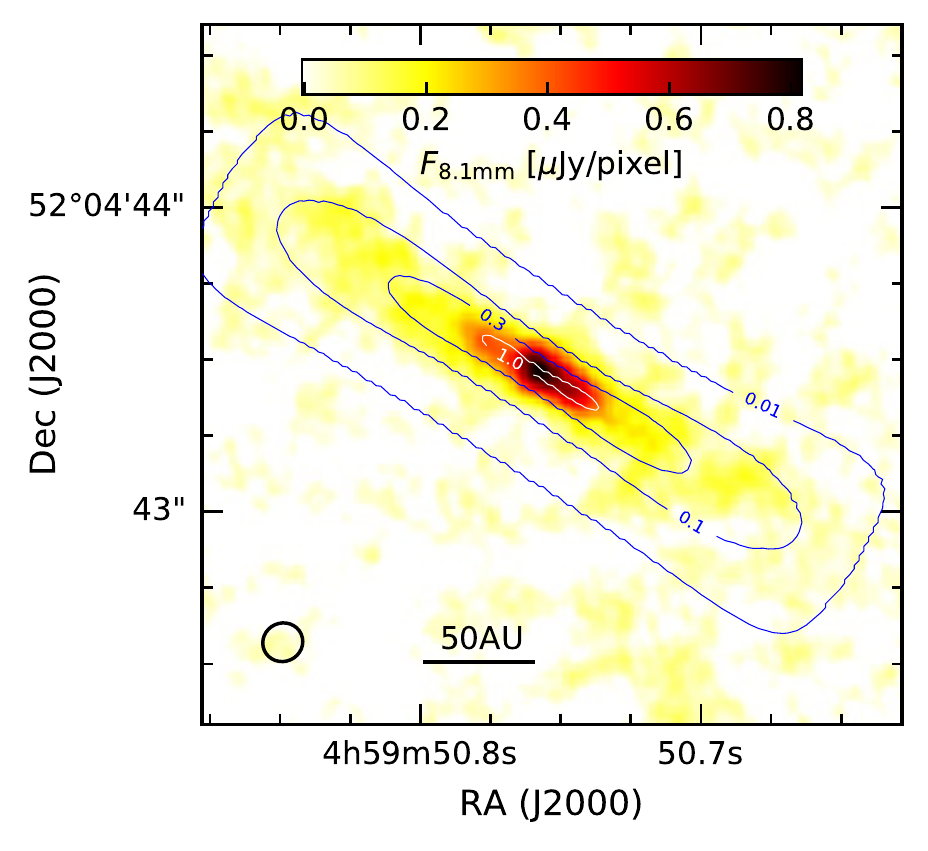}
\caption{Modelled optical depth contours ($\tau_{\rm1.3mm}$, $\tau_{\rm2.9mm}$, $\tau_{\rm8.1mm}$), overlaid on the observed intensity maps at 1.3, 2.9, and 8.1\,mm, respectively. The observed intensity maps are identical to those shown in the leftmost panels of Fig.\,\ref{fig_obs_sim_2d}, and show the dust emission convolved with the indicated beam sizes, while the optical depth contours are not convolved. The optical depth value is indicated at each corresponding contour. The zoomed-in optical depth distributions are displayed in Fig.\,\ref{fig_tau_beta}.}
\label{fig_tau}
\end{figure*}

\begin{figure*}[htp]
\centering
\includegraphics[width=0.33\textwidth, angle=0]{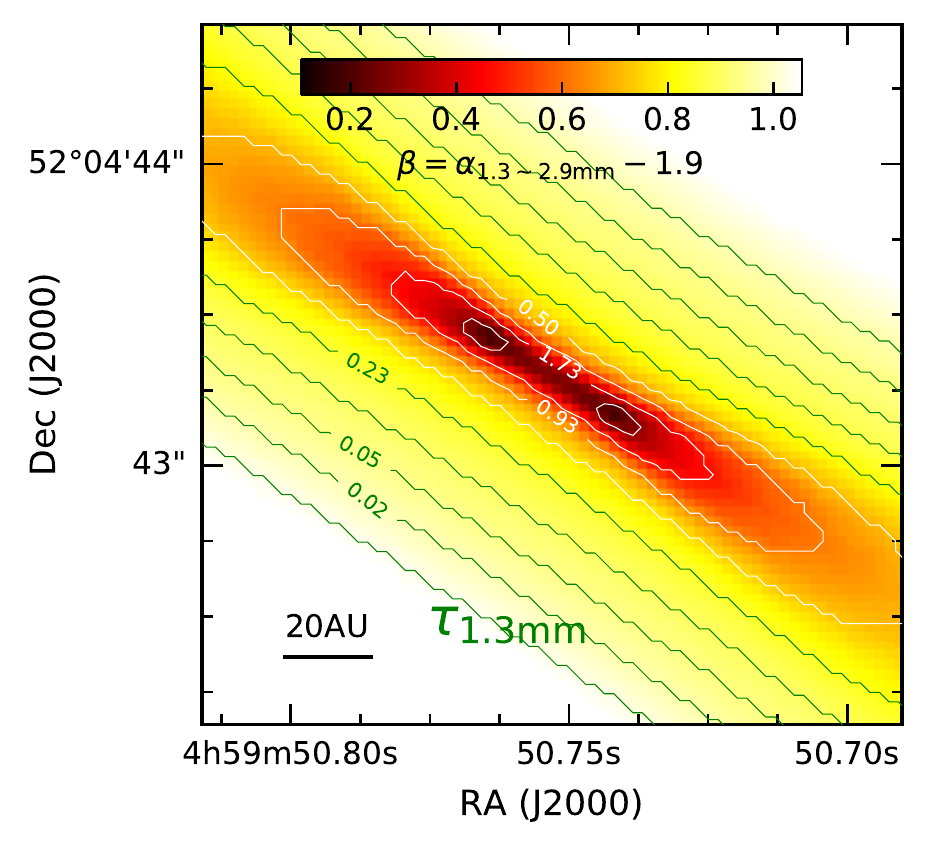} 
\includegraphics[width=0.33\textwidth, angle=0]{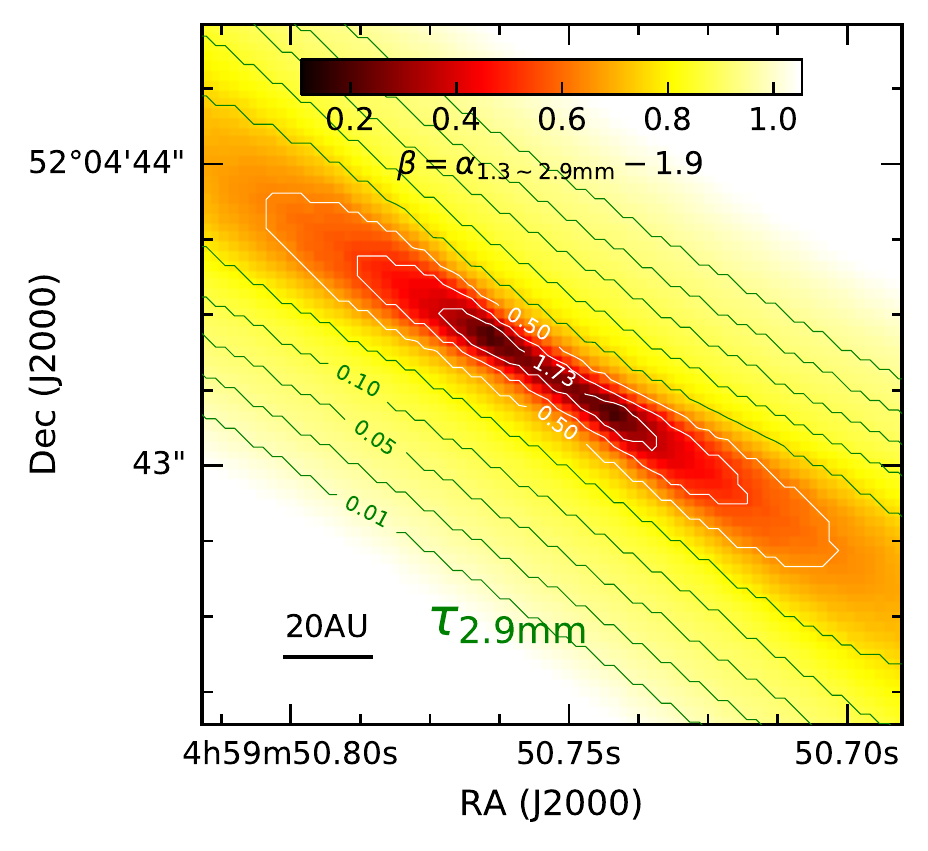}
\includegraphics[width=0.33\textwidth, angle=0]{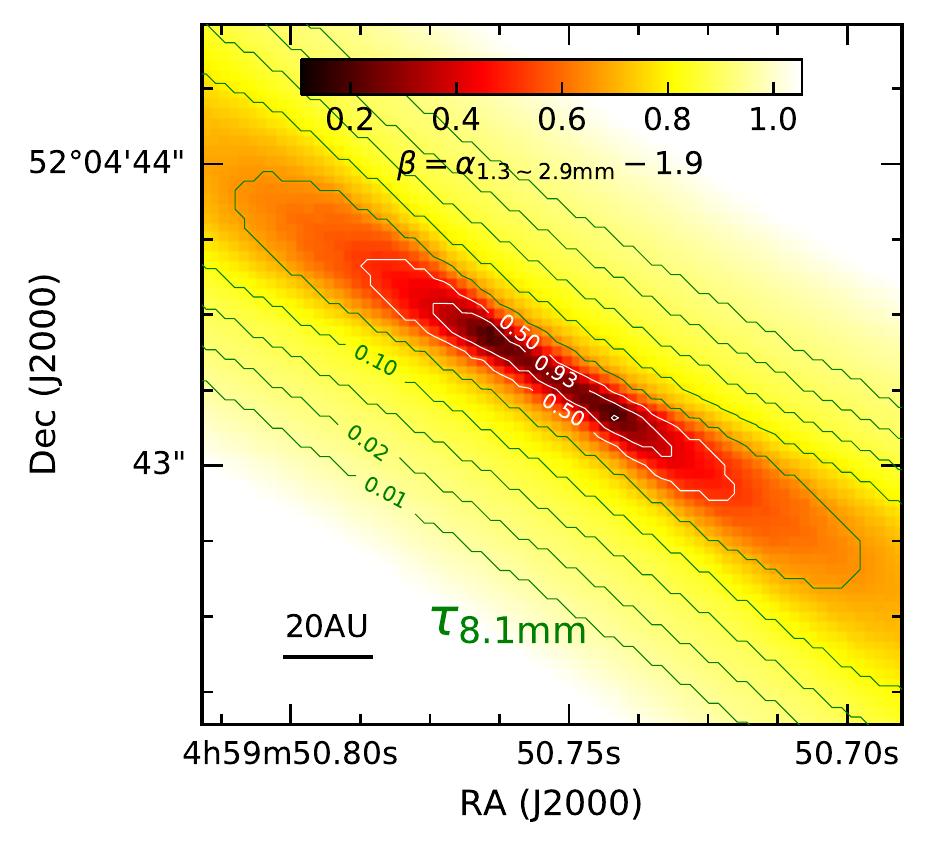}
\caption{Zoomed-in modelled optical depth contours at 1.3, 2.9, and 8.1\,mm, overlaid on the (also unconvolved) colour map of the millimeter opacity slope, $\beta_{\rm mm}$, derived from the best-fit model maps (cf. Figs.\,\ref{fig_obs_sim_2d} and \ref{fig_beta}). The optical depth values are indicated at each corresponding contour. The pixel size for all three maps is $0.016''\times0.016''$.}
\label{fig_tau_beta}
\end{figure*}

Figure\,\ref{fig_temperature} displays the temperature structure as a function of radius $r$ and polar angle $\theta$ in spherical coordinates. Figure\,\ref{fig_temperature_1d} shows the radial temperature distribution along the midplane of the disk and envelope. The envelope temperature profile generally follows a power-law, both inside the inner hole of the disk and outside the disk. Snow-lines of volatile species play an important role in planet formation theories \citep[e.g.,][]{Pontoppidan2014}. Assuming a water ice sublimation temperature range of 150$-$170\,K \citep[e.g.,][]{Podolak2004,Liu2017}, the water snowline in the disk midplane is located at a radius range of $\sim$4.1--5.4\,au, much smaller than the inner radius ($r_{\rm in}=16$\,au) of the disk (see Figs.\,\ref{fig_temperature} and \ref{fig_temperature_1d}). The temperature in the disk is $20\la T_{\disk}\la90$\,K, which is much lower than the water ice sublimation temperature. Close to the outer edge of the disk, where, by chance, the CO snow line ($\sim$20\,K) is located, the radial temperature profile turns flat and gradually becomes steeper when going further inward in the disk. CO freezes out at about 20\,K and the disk as a whole does not get below 20\,K. This is further evidence that CB\,26 is still a very young embedded disk, which tends to be warmer than more evolved protoplanetary disks \citep{Burke2010,Hoff2018a,Hoff2018b}.


\subsection{Mass}
\label{sect_mass}

Our best-fit model suggests dust masses in the disk and envelope of \mbox{$M_\disk^{\rm dust}=7.6\times10^{-4}$\,\msol} and \mbox{$M_\env^{\rm dust}=3.3\times10^{-4}$\,\msol}, respectively. Assuming a constant gas-to-dust mass ratio of 100 \citep{Glauser2008}, their gas masses are \mbox{$M_\disk^{\rm gas}=7.6\times10^{-2}$\,\msol} and \mbox{$M_\env^{\rm gas}=3.3\times10^{-2}$\,\msol}, respectively. 
Uncertainties are difficult to quantify because they largely depend on the absolute value of the dust absorption coefficient and on the unconstrained gas-to-dust mass ratio. Furthermore, the total extent of the envelope is also poorly constrained and the choice of our cut-off radius at 900\,au (Sect.\,\ref{sect_env}) is arbitrary. Based on single-dish mm and {\it Herschel} FIR data, \citet{Launhardt2013} derive an envelope radius and total mass of $1\times10^4$\,au and $\approx$0.22\,\msol\footnote{\mbox{0.3\,\msol} minus disk mass}, respectively.
The mass of the disk thus amounts to about 14\% of the central star mass \citep[\mbox{$M_{\star}=0.55$\,\msol;}][]{Launhardt2020}, which is consistent with hydrodynamic simulations of marginally self-gravitating disks \citep{Nelson2000}. Thus, the CB\,26 disk stands out as one of the most extended and massive Class\,I disks \citep[e.g.,][]{Tobin2020,Tychoniec2020}. \citet{Hogerheijde2001} showed that rotationally supported disks grow as long as they are still embedded and accrete from the envelope, and start decreasing in size when the envelope is dispersed. This implies that such disks reach their maximum size at the transition from the fully embedded to the T Tauri phase. With CB\,26 we have caught a disk just at this transition stage when it reaches the maximum disk size.


\subsection{Optical depth}
\label{sect_tau}

\begin{figure}[htp]
\centering
\includegraphics[width=0.45\textwidth, angle=0]{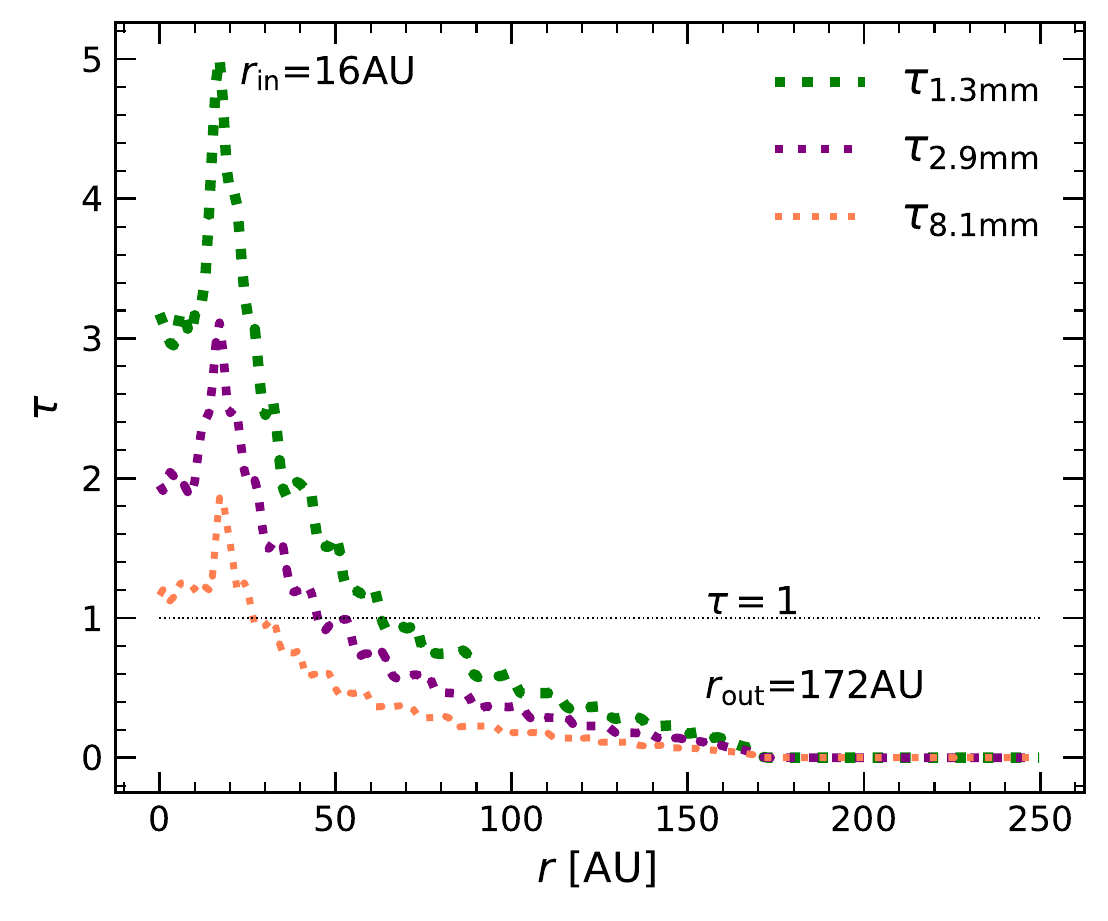} 
\caption{Modelled dust optical depth distribution along the major radial direction at 1.3, 2.9, and 8.1\,mm in Fig.\,\ref{fig_tau_beta}.}
\label{fig_tau_radius}
\end{figure}

Figure\,\ref{fig_tau} displays the modelled optical depth contours of the dust emission, overlaid on the observed intensity maps at 1.3, 2.9, and 8.1\,mm, respectively. Figure\,\ref{fig_tau_beta} shows the zoomed-in optical depth distributions, overlaid on . Figure\,\ref{fig_tau_radius} shows the optical depth distributions at 1.3, 2.9, and 8.1\,mm along the major radial direction. These optical depth distributions and profiles demonstrate that the dust emission along the disk midplane is only moderately optically thick in the innermost parts of the disk ($r_{\rm 1.3mm}\la 65$\,au, $r_{\rm 2.9mm}\la50$\,au, and $r_{\rm 8.1mm}\la30$\,au), while most parts of the outer disk are optically thin at all three wavelengths.


\subsection{Grain size distribution}
\label{sect_grainsize}

For both the envelope and the disk we adopted a power-law grain size distribution with exponent $-$3.5 (eq.\,\ref{equ_grainsize}), a minimum grain size of \mbox{$a_{\rm min}=10\nm$}, and let the maximum grain size vary in the modelling (Sect.\,\ref{sect_grain-size-distr}). As indicated in Table\,\ref{tab_model}, we derive a best-fit maximum grain size for the envelope emission of \mbox{$a_{\rm max}^{\rm env}\approx0.1^{+0.2}_{-0.1}$\,mm}, which is already significantly larger than upper grain sizes if ISM dust given in the literature. Our best-fit models for the maximum grain size in the disk indicate \mbox{$a_{\rm max}^{\rm disk}\approx5^{+4}_{-2}$\,cm}. Hence, the dust emission from the disk is incompatible with both ISM-type dust and the grain size distribution in the envelope, but requires the existence of already pebble-sized bodies with diameters of about 10\,cm. In Sect.\,\ref{sect_growth}, we discuss the robustness and implication of this finding.


\section{Analysis and Discussion}
\label{sect_discussion}


\subsection{Free-free emission at 8.1\,mm?}
\label{sect_free-free}

The physics behind the ionization around high-mass stars is reasonably well understood as arising from photoionization due to the strong UV flux from such objects. Ionization around low-mass stars is much less-understood. A few mechanisms have been proposed for low-mass stars to produce the ionization responsible for the detected free-free emission \citep{Anglada1998,Scaife2012}. The centimeter-wavelength emission of young stars is often dominated by hot plasma in the system, either gyrosynchrotron emission associated with chromospheric activity or thermal bremsstrahlung emission from partially ionized winds \citep{Wilner2005}. 

In CB\,26, the sharp central peak in the 8.1\,mm map cannot be fitted with only dust emission in the simulation. Instead, the observed intensity distribution requires the contribution from a compact non-thermal source in addition to the thermal dust emission from the disk (see Figs.\,\ref{fig_obs_sim_2d} and \ref{fig_obs_sim_1d}). Assuming that this compact non-thermal source is free-free emission and that the fluxes at 4 and 6.4\,cm are dominated by this free-free emission, the estimated spectral index for the non-dust emission would be $\approx$0.13. This indicates that the ionized wind from star is quite inhomogeneous \citep{Condon2016}. One can also extrapolate that the free-free contribution at 8.1\,mm is $\approx$77\,$\mu$Jy, which accounts for $\approx$7\% of the observed total 8.1\,mm flux (see bottom panel of Fig.\,\ref{fig_obs_sim_1d}). The residual of the sharp peak between observation and simulated dust emission amounts to about $\approx$110\,$\mu$Jy or 10\% of the observed total flux at 8.1\,mm. Hence, the total flux of the sharp central peak at 8.1\,mm is comparable to the extrapolated free-free contribution at this wavelength. We can therefore safely assume that the sharp peak is mainly produced by free-free emission, while the wide wings in the 8.1\,mm map are produced by dust emission. 

Since CB\,26 is driving a collimated bipolar outflow \citep{Launhardt2009}, a jet-origin (shock ionization) of this free-free emission is the most likely scenario. An alternative explanation could be that the central peak at 8.1\,mm is tracing a photoevaporating region in the circumstellar disk \citep{Pascucci2012,Rodriguez2014}. This photoevaporation region must be much smaller than the observed dust disk and could well occupy and cause the inner whole that is observed at millimeter wavelengths (Sect.\,\ref{sect_hole}).

While we can explain the central peak in the 8.1\,mm emission map (Fig.\,\ref{fig_obs_sim_1d}) with compact free-free emission (Sect.\,\ref{sect_free-free}), there is still significant extended residual emission at \mbox{$r\approx50-65$\,au} when only the best-fit dust emission model of the disk and an unresolved free-free emission region are taken into account (Figs.\,\ref{fig_obs_sim_2d} and \ref{fig_obs_sim_1d}). We discuss this residual excess emission in Sect.\,\ref{sect_growth}.


\subsection{The inner hole}
\label{sect_hole}

The observed 1.3\,mm intensity distribution along the major axis of the disk, shown in Fig.\,\ref{fig_obs_sim_1d}, exhibits a suspicious plateau at the center, similar but less prominent than that in the OVRO-only map used by \citet{Sauter2009}. This plateau is less prominent or not visible at all in the 1.1, 2.9, and 8.1\,mm maps. At 1.1 and 2.9\,mm, this may be related to the lower angular resolution of these maps (Tab.\,\ref{tab_beam}). The 8.1\,mm map has a higher angular resolution, but the compact free-free emission (Sect.\,\ref{sect_free-free}) may mask a possible plateau or dip. The plateau in the 1.3\,mm map may be caused by an inner hole in the disk or could be the result of the (moderate) optical thickness of the millimeter dust emission from the innermost parts of the disk (Fig.\,\ref{fig_tau_radius}), such that enough flux from the (warmer) inner disk is obscured.

In our modelling, pure optical depth effects in a disk without inner hole could clearly not reproduce the observed dust emission profiles. Instead, an inner whole was needed. In our model, this inner whole is not completely dust-free, but is still filled with the dust from the envelope model (Sect.\,\ref{sect_env}). The size of this inner hole is poorly constrained by our models. We derive a best-fit radius of \mbox{$r_{\rm in} = 16^{+37}_{-8}$\,au} (Tab.\,\ref{tab_model}). The observed plateau at 1.3\,mm could be reproduced in the simulations only if we force the size of the inner hole to be $r_{\rm in} \ga 30$\,au, which is still within the range of uncertainties given above. In this case, the plateau would still not be visible at 1.1\,mm and 2.9\,mm (due to the large beam size), like in the observed maps.


\subsection{Grain growth}
\label{sect_growth}

\begin{figure}[htp]
\centering
\includegraphics[width=0.45\textwidth, angle=0]{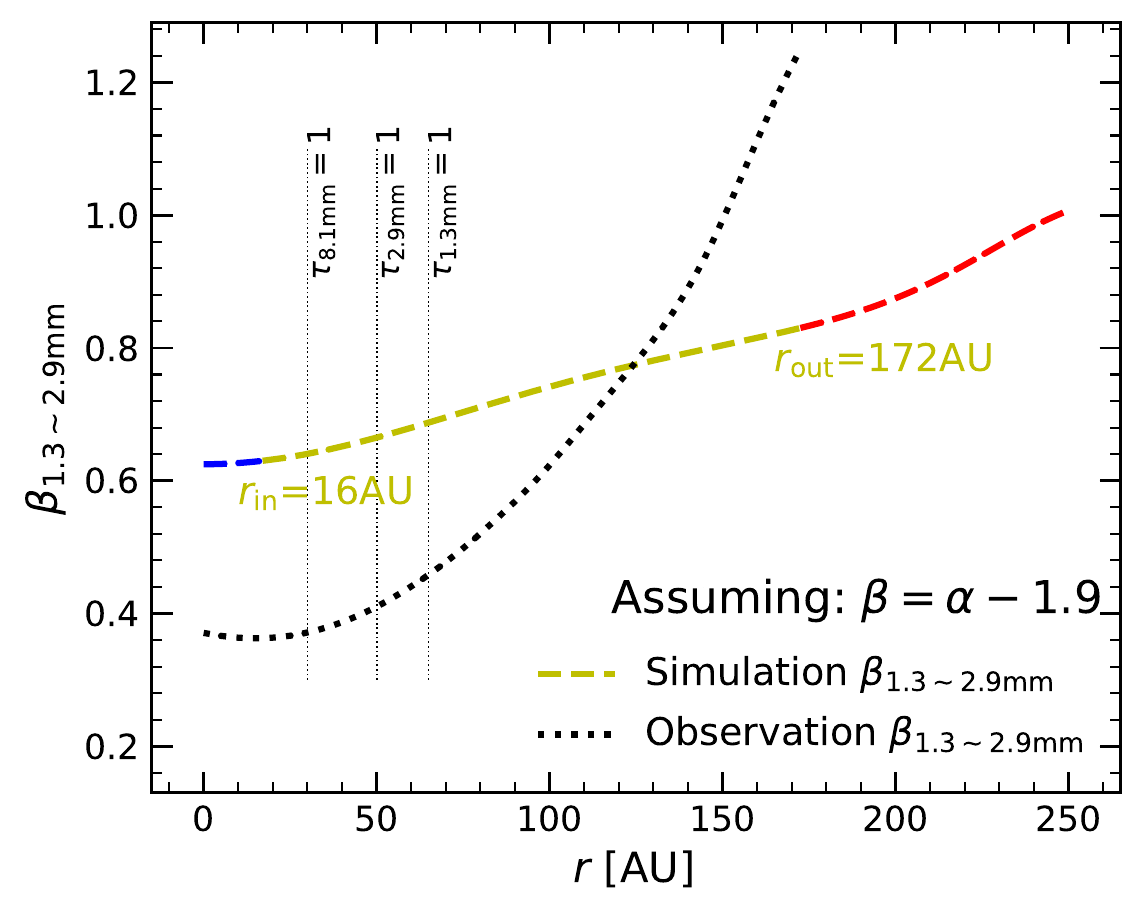}
\includegraphics[width=0.45\textwidth, angle=0]{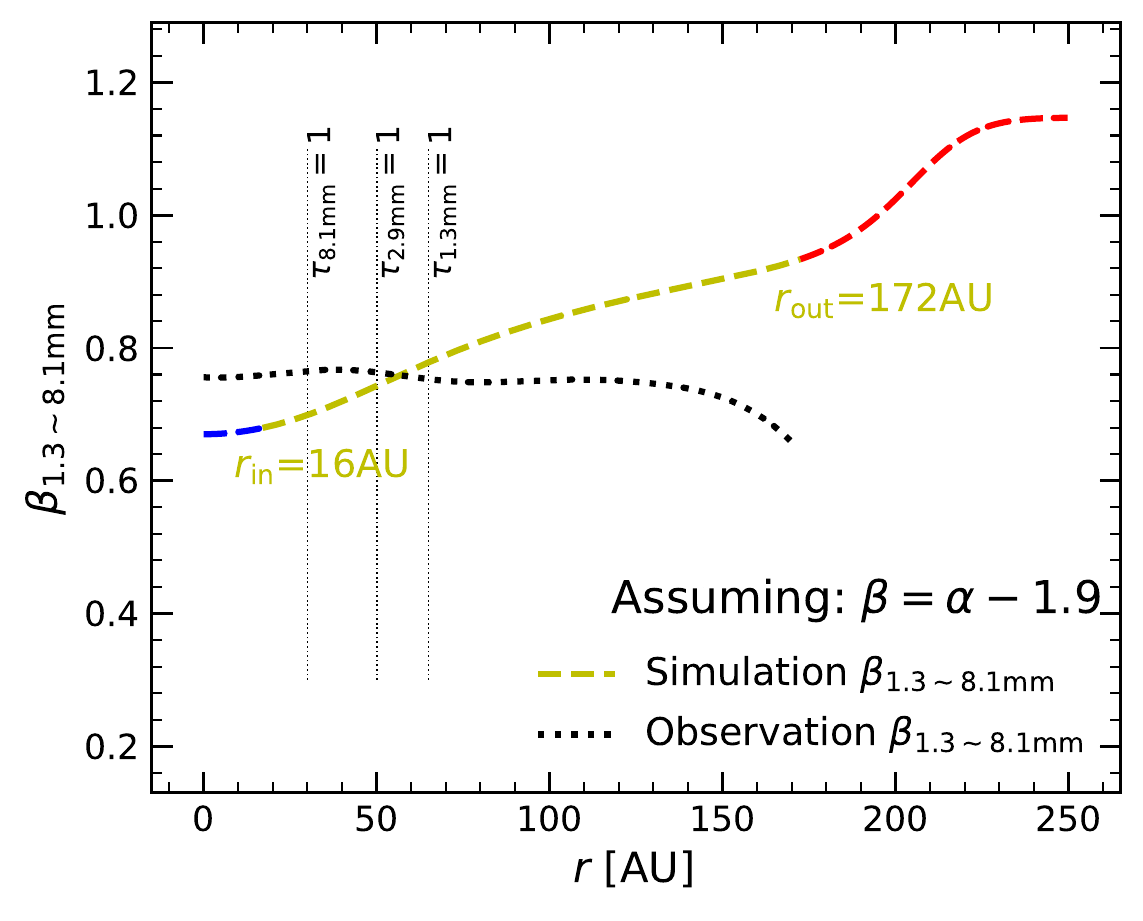}
\caption{Millimeter absorption opacity slopes $\beta_{\rm mm}$ along the major axis of the disk in the observations and the best-fit model. The slopes $\beta_{\rm 1.3\sim2.9mm}$ and $\beta_{\rm 1.3\sim8.1mm}$ are derived from spectral indexes $\alpha_{\rm 1.3\sim2.9mm}$ between 1.3 and 2.9\,mm, and $\alpha_{\rm 1.3\sim8.1mm}$ between 1.3 and 8.1\,mm, respectively (see text). All data have been convolved to the angular resolution of the CARMA 2.9\,mm map (Tab.\,\ref{tab_beam}). The contribution from free-free emission at 8.1\,mm has been removed (see Sect.\,\ref{sect_fit} and Fig.\,\ref{fig_obs_sim_1d}).}
\label{fig_beta}
\end{figure}

Our best-fit radiative transfer model, which has $a_{\rm max}$ of the grains size distribution as a free parameter (Tab.\,\ref{tab_model}), and the uncertainty ranges indicate that the dust grains in the disk have already grown to pebbles with diameters of the order 10\,cm (\mbox{$a_{\rm max}^{\rm disk}\approx5^{+4}_{-2}$\,cm}, Tab.\,\ref{tab_model}). Even the dust in the envelope contains at least submm-sized grains (\mbox{$a_{\rm max}^{\rm env}\approx0.11^{+0.22}_{-0.07}$\,mm}), which is several orders of magnitude larger than the upper end of the grain size distribution in the ISM \citep[e.g.,][]{Weingartner2001}. If we use maximum grain sizes $a_{\rm max}^{\rm disk} = 2.5\mum$ in the disk and $a_{\rm max}^{\rm env} = 0.25\mum$ in the envelope, as in \citet{Sauter2009}, we cannot reproduce the SED, even after testing a large parameter space. The reason for this discrepancy with \citet{Sauter2009} is most likely manifold. Besides a slightly different treatment of the envelope, we could in our new study include the {\it Herschel} FIR measurements and, more importantly, the new 8.1\,mm VLA data, which provide much more robust constraints on the mm dust emissivity than the 1.3\,mm and 3\,mm data only. As we show in Appendix\,\ref{app_withsma} and discuss in Sect.\,\ref{sect_chi2}, our result is not affected by in- or excluding the 1.1\,mm SMA map, and can thus be considered robust.

The best-fit modelled optical depth distributions, displayed in Fig.\,\ref{fig_tau_radius} (see also Sect.\,\ref{sect_tau}), shows that the thermal emission at all mm wavelengths is optically thin ($\tau<1$) in the outer disk ($r\gtrsim60$\,au), and becomes only moderately optically thick ($\tau_{\rm1.3mm}<5.0$, $\tau_{\rm2.9mm}<3.1$, $\tau_{\rm8.1mm}<1.9$) in the innermost parts of the disk at $r\lesssim50$\,au, and even there only close to the disk mid-plane (Figs.\,\ref{fig_tau} and \ref{fig_tau_beta}). Thus, we can exclude that a not optimally modelled optical depth distribution could have significantly affected our results. As explained in Sect.\,\ref{sect_scattering}, we have also taken into account scattering in our modelling (Fig.\,\ref{fig_dustkappa}), such that the two main simplifications that could lead to an overestimation of $a_{\rm max}$ are avoided here \citep[e.g.,][]{Carrasco2019}.

Even if we ignore scattering (Sect.\,\ref{sect_scattering}), we can still reproduce the mm SED and maps of CB\,26 with an only slightly differing best-fit value of $a_{\rm max}$. The reason why scattering makes a much smaller difference for CB\,26 as compared to, e.g., the ALMA-based study of HL\,Tau presented recently by \citet{Carrasco2019}, is manifold. First, the peak optical depth in our  CB\,26 data is much smaller than in the ALMA data of HL\,Tau. This is related to both the higher angular resolution of the ALMA data as compared to our data (both sources are located at approximately the same distance), and to the fact that CB\,26 is seen edge-on, while HL\,Tau is seen nearly pole-on. Due to both the different beam sizes and the different projections, small-scale structures with locally increased optical depths are much better resolved in HL\,Tau than in CB\,26. This in turn could imply that possible substructures (rings, spirals, clumps) with increased optical depths and a potentially differing grain size distribution might be hidden, since unresolved and not modelled in our CB\,26 data (our disk model does not include small-scale substructures since there are no such observational constraints, see Sect.\,\ref{sect_disk}). On the other hand, such hypothesized over-dense small-scale substructures in the CB\,26 disk would occupy only a tiny areal fraction and contribute only very little to the observed emission, since they are expected to be confined to the disk midplane, which is seen edge-on. Hence, even if present, such small-scale structures would not significantly affect our results for the bulk of the dust in the CB\,26 disk.

In our model, grain sizes are well-mixed throughout the disk and possible vertical settling of the largest grains has not been taken into account. To verify if ignoring vertical dust settling could have affected our conclusions on grain growth in the CB\,26 disk, we also perform a test in which the vertical scale height ($h$; see also eq.\,\ref{equ_scalehight}) is a function of the grain size, $a$:
\begin{equation}
h(a) = h(a_{\rm min}) \left(\frac{a}{a_{\rm min}}\right)^{\xi}.
\label{equ_scalehight_vertical}
\end{equation}
Here, $h(a_{\rm min})$ is the scale height for the smallest dust grains, while the parameter $\xi$ quantifies the degree of dust settling. We fix $\xi=-0.1$, which is typically found for other systems, for instance the IM Lup disk \citep{Pinte2008}. We find that the millimetre fluxes of the settled disk are overall slightly lower than those of the well-mixed model. This is due to the fact that large dust grains are concentrated close to the midplane where the temperature is low. To compensate the flux deficit, a higher total dust mass in the disk is required to fit the observation, which in turn results in a slightly larger optical depth. Nevertheless, as long as large dust grains are only moderately depleted, but still present at larger scale heights, the slope of the millimeter SED remains nearly unchanged. Therefore, we conclude that dust settling does not have a significant impact on the derived maximum grain size.

However, as we show in Sect.\,\ref{sect_free-free} and Figs.\,\ref{fig_obs_sim_2d} and \ref{fig_obs_sim_1d}, our best-fit model with \mbox{$a_{\rm max}=5$\,cm} throughout the disk and a compact (unresolved) central region with free-free emission still leaves significant extended 8.1\,mm residual emission at \mbox{$r\approx50-65$\,au}. Since the 8.1\,mm thermal dust emission at these radii is clearly optically thin (Fig.\,\ref{fig_tau_radius}), and the moderate optical thickness of the emission at shorter wavelengths has been taken into account in our modelling, this residual emission could hint at a radial change in the dust grain size distribution, i.e., the presence of even larger grains in the inner $\approx$65\,au of the disk. Because of the uncertainties involved at this level of detail and the missing observational constraints on size scales <50\,au (the projected beam size at 8.1\,mm is $\approx$18\,au, while it is only $\approx$52\,au at 1.3\,mm; see Tab.\,\ref{tab_beam}), we do not involve a model with a step function or radial gradient in $a_{\rm max}(r)$, and thus do not attempt to quantify this possible further grain growth in the inner disk.

Since the dust emission is optically thin throughout most of the disk, a large value of $a_{\rm max}$, i.e. the presence of dust grains that are significantly larger than the observing wavelength, should become directly evident in the spectral index of the mm dust emission, $\alpha_{\rm mm}$, which relates for optically thin emission to the spectral index of the dust absorption coefficient as \mbox{$\beta_{\rm mm}=\alpha_{\rm mm}-\alpha_{\rm P}$}, where $\alpha_{\rm P}=d{\rm lg}B(\nu,T)/d{\rm lg}\nu$ is the spectral slope of the Planck function. With a dust temperature $T\approx 20\ldots 100$\,K in disk (Fig.\,\ref{fig_temperature}), the spectral index will be $\alpha_{\rm p}=1.80\ldots 1.96$ between 1.3 and 2.9\,mm. Considering that the majority of the disk (Fig.\,\ref{fig_temperature_1d}) has a dust temperature of around 40\,K, we could reasonably adopt $\alpha_{\rm P}=1.9$ throughout the disk, thus we assume \mbox{$\beta_{\rm mm}=\alpha_{\rm mm}-1.9$} in this work.
Indeed, we  derive a millimeter spectral index of the observed integrated fluxes at 1.3 and 2.9\,mm of \mbox{$\alpha_{\rm 1.3\sim2.9mm}=2.63\pm0.40$}, which for optically thin emission corresponds to a mean absorption opacity slope of \mbox{$\beta_{\rm 1.3\sim2.9mm}=0.73\pm0.40$}. For the corresponding 1.3 to 8.1\,mm slope we obtain \mbox{$\alpha_{\rm 1.3\sim8.1mm}=2.56\pm0.30$}, which is consistent with $\alpha_{\rm 1.3\sim2.9mm}$. 

Furthermore, Fig.\,\ref{fig_beta} shows the spatially resolved millimeter opacity slope profiles, derived under the optically thin assumption from the spectral index slopes, along the major axis of the disk in the observations and the best-fit model. Both are convolved at all wavelengths to the beam size of the 2.9\,mm map (Table\,\ref{tab_beam}).
Comparing the observations directly, we derive \mbox{$0.36\lesssim\beta_{\rm 1.3\sim2.9mm}\lesssim1.24$} with a relatively strong positive radial gradient (Fig.\,\ref{fig_beta}). Comparison of the 1.3\,mm map with the longer wavelength map at 8.1\,mm results in an approximately radially constant slope of \mbox{$\beta_{\rm 1.3\sim8.1mm}\approx0.73\pm0.05$}. Since already a slight mismatch in the effective beam sizes or a slight pointing offset between the maps at different wavelength can result in such a radial gradient, we do not try to interpret these gradients here, but look at the simulated best-fit maps instead, which do not have these problems. 

Comparing the convolved simulated best-fit maps, we derive \mbox{$0.62\lesssim\beta_{\rm 1.3\sim2.9mm}\lesssim0.82$} and \mbox{$0.67\lesssim\beta_{\rm 1.3\sim8.1mm}\lesssim0.92$}, respectively, with a slight positive radial gradient in both cases (Fig.\,\ref{fig_beta}). This  positive radial gradient is expected and is mainly caused by the relative contribution from the envelope (with smaller dust grains and larger $\beta_{\rm mm}$), which increases with increasing impact parameter. Also contributing, but to a lesser degree, is the non-negligible optical depth, which increases towards smaller impact parameters (Fig.\,\ref{fig_tau_radius}), and the radial temperature gradient (Fig.\,\ref{fig_temperature}). Altogether, the spectral slope of the integrated fluxes, as well as the spatially resolved radial profiles of the spectral slopes derived from both the observed and the best-fit modelled maps all consistently suggest a mean mm absorption opacity slope of \mbox{$\beta_{\rm mm}\approx0.7\pm0.4$} for the outer disk ($r\gtrsim50$\,au) of CB\,26.

Such a small value of $\beta_{\rm mm}<1$ is understood to indicate that the dust particles have grown into large grains of at least millimeter size or even larger \citep[e.g.,][]{Testi2003,Rodmann2006,Pinte2008,Tobin2013,Tazzari2016}. We want to stress here that the mm spectral slope in CB\,26 can only be related so directly to the maximum grain size because our models have shown that the emission is optically thin.  
\citet{Ricci2010a} derive a similar value for the mean absorption opacity slope in $\rho$-Ophiuchi protoplanetary disks, in which the grains are also thought to have grown to millimeter or centimeter sizes.

The presence of cm-sized dust grains in the about 1\,Myr old CB\,26 disk is consistent with models of grain growth and radial drift in protoplanetary disks by, e.g., \citet{Birnstiel2010}. Moreover, our finding indicates that solids grow rapidly already during the first million years in a protostellar disk, making it thus possible that the Class\,II (transitional) disks are already seeded with large particles, making it much more likely that they already contain large bodies and perhaps even planetesimals \citep[e.g.,][]{johansen2014}. Furthermore, the dust masses of Class\,II disks inferred from mm surveys \citep[e.g.,][]{Andrews2013,Ansdell2016,Pascucci2016} could partially reflect this growth of solids to large sizes that are no longer emitting efficiently.


\subsection{Toomre stability}
\label{sect_instability}

\begin{figure}[htp]
\centering
\includegraphics[width=0.45\textwidth, angle=0]{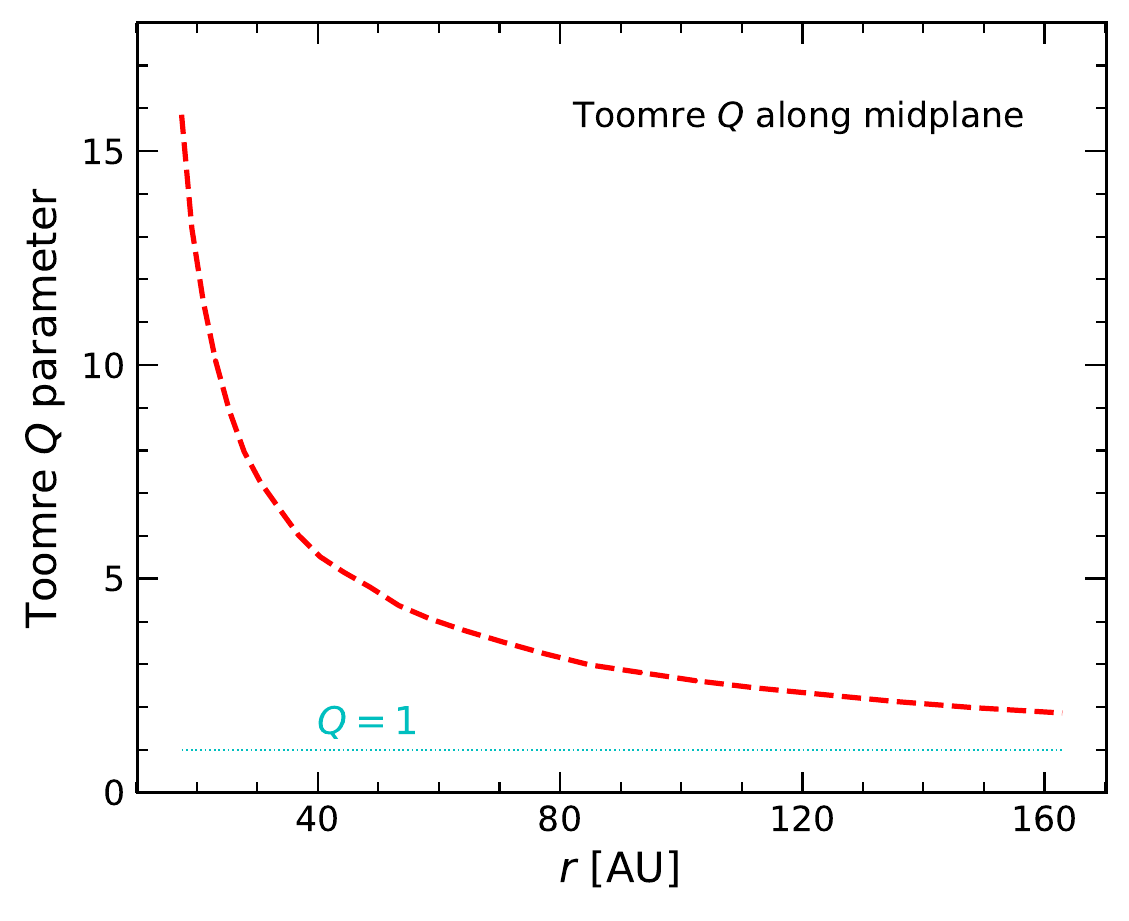}
\caption{Toomre $Q$ parameter as a function of radius $r$ along the disk midplane. $Q=1$ is indicated with a dotted line.}
\label{fig_toomre}
\end{figure}

The gravitational stability of the disk can be tested using the Toomre criterion \citep{Toomre1964}, which describes when a disk becomes unstable to gravitational collapse. The Toomre $Q$ parameter is defined as
\begin{equation}
Q=\frac{c_{\rm s}\Omega_{\rm K}}{\pi G \Sigma_{\rm gas}},
\end{equation}
where \mbox{$c_{\rm s} = \sqrt{kT/(\mu m_{\rm p})}$} is the local sound speed, \mbox{$\Omega_{\rm K}=\sqrt{GM_{\star}/r^3}$} is the angular velocity for Keplerian rotation \citep{Knigge1999}, and $\Sigma_{\rm gas}$ stands for the surface density of the gas in the disk, which can be derived from the surface density of the dust assuming a constant gas-to-dust mass ratio of 100 throughout the disk for simplicity \citep{Glauser2008}. The corresponding parameters are listed in Table\,\ref{tab_model}. In terms of $Q$, a disk is unstable to its own self-gravity if $Q < 1$, and stable if $Q > 1$. Figure \ref{fig_toomre} displays the Toomre $Q$ parameter as a function of radius in the CB\,26 disk. It clearly shows that $Q$ is greater than unity for the entire disk ($1.8\la Q \la 16.4$), indicating that the disk is gravitationally stable. However, in contrast to the inner disk, the outer parts of the disk at $r\ga 80$\,au have $Q$ values close to unity \mbox{($1.8\la Q \la 3.5$)}, suggesting it is approaching marginal instability. In numerical simulations of rapid accretion, \citet{Kratter2010} found that a marginally unstable outer disk will not fragment, but may generate spiral arms \citep[see also, e.g.,][]{Rice2003,Stamatellos2008,Dong2018}. Unfortunately, this cannot be verified directly observationally in CB\,26 due to the extreme edge-on orientation.


\section{Summary and conclusions}
\label{sect_summary}

The Bok globule CB\,26 harbours a young ($\approx$1\,Myr) low-mass ($\approx$0.55\,\msol) star that is surrounded by a massive circumstellar disk seen nearly edge-on, which is in turn still embedded in a thin remnant envelope. Combining the well-sampled SED between 0.9$\mum$ and 6.4\,cm, high angular resolution millimeter continuum observations at 1.1, 1.3, 2.9, and 8.1\,mm, and using the radiative transfer code \texttt{RADMC-3D} \citep{Dullemond2012}, we conduct a detailed modelling of the dust emission from the CB\,26 circumstellar disk. The models are fitted simultaneously to the complete SED and to the image plane data of the interferometric dust continuum emission maps at 1.3, 2.9, and 8.1\,mm. 

From the best-fit models to the thermal dust emission, we infer an outer radius of the disk of \mbox{$r_{\rm out}=172\pm22$\,au}, and a size of the inner hole of \mbox{$r_{\rm in}=16^{+37}_{-8}$\,au}. Assuming a constant gas-to-dust mass ratio of 100, we derive a total gas mass in the disk of \mbox{$M_{\rm disk}=7.6\times10^{-2}$\,\msol}. The mass of the disk amounts to $\approx$14\% of the mass of the central star. The mass of the surrounding envelope remains largely unconstrained since our model takes into account only the innermost 900\,au \mbox{($3.3\times10^{-2}$\,\msol$<M_{\rm env}\lesssim0.22$\,\msol)}. The other best-fit model parameters are listed in Table\,\ref{tab_model}. Thus the disk around CB\,26 stands out as one of the most extended and massive Class\,I disks.

We find that the central part of the disk contains an unresolved compact region of free-free emission, which significantly contributes to the 8.1\,mm flux and completely dominates the cm emission. We cannot constrain whether this free-free emission is related to a jet (shock ionization) or originates from a photoevaporation region in the inner hole of the disk. For our modelling of the thermal dust emission, we treat this free-free emission region as point-like, extrapolate its contribution to the 8.1\,mm emission from the spectral slope of the 4.0 and 6.4\,cm fluxes, subtract it from the 8.1\,mm map, and exclude the contaminated central region in the 8.1\,mm map from the model fitting.

We find that the thermal dust emission from the outer disk \mbox{($r\gtrsim60$\,au)} is optically thin ($\tau<<1$) at all mm wavelengths, while the emission from the midplane of the inner disk ($r\lesssim50$\,au) becomes moderately optically thick (\mbox{$\tau_{\rm1.3mm}<5.0$}, \mbox{$\tau_{\rm2.9mm}<3.1$}, \mbox{$\tau_{\rm8.1mm}<1.9$}). Our best-fit radiative transfer models, which assume a grain size distribution with power-law exponent $-3.5$ and \mbox{$a_{\rm min}=5$\nm}, indicate that the dust grains in the disk have already grown to pebbles with diameters of the order 10\,cm. Even the dust in the envelope contains already at least submm-sized grains, i.e., much larger than what is considered the upper end of the grain size distribution in the ISM. 

Our best-fit model still leaves significant extended 8.1\,mm residual emission at \mbox{$r\approx50-65$\,au}, which we interpret as a radial change in the grain size distribution and the presence of even larger particles in the inner disk. However, we are not able to reliably quantify their size. 

Since our detailed models show that the mm dust emission from the outer disk is completely optically thin, and the contribution from moderately optically thick emission from the midplane in the inner disk is very small, we can directly relate the spectral slope of the mm dust emission, $\alpha_{\rm mm}$, to the spectral index of the dust absorption coefficient, $\beta_{\rm mm}$, and to the size of the largest dust grains, $a_{\rm max}$, when these are of the same order or larger than the observing wavelength. Indeed, we derive a mean mm absorption opacity slope of \mbox{$\beta_{\rm mm}\approx0.7\pm0.4$} for the entire outer disk of CB\,26, which is consistent with the presence of mm-to-cm-sized dust particles.

Altogether, our robust finding of the presence of at least cm-sized bodies in the CB\,26 disk indicates that solids grow rapidly already during the first million years in a protostellar disk, making it thus possible that the slightly more evolved Class\,II (transitional) disks are already seeded with large particles. This in turn makes it likely that these disks may already contain larger bodies and perhaps even planetesimals.

Last but not least, we find the inner disk of CB\,26 ($r\lesssim80$\au) is gravitationally stable, while the outer disk may be approaching marginal instability and thus be prone to generating spiral arms. Unfortunately, this cannot be verified observationally due to the extreme edge-on orientation of CB\,26.

\begin{acknowledgements}
We thank the anonymous referee for constructive comments that improved our study. This work is supported by the National Natural Science Foundation of China Nos.\,11703040, 11743007, 11503087, 11973090, and the Natural Science Foundation of Jiangsu Province of China No.\,BK20181513. C.-P.Z. acknowledges supports from the MPG-CAS Joint Doctoral Promotion Program (DPP), the China Scholarship Council (CSC) in Germany as a postdoctoral researcher, the NAOC Nebula Talents Program, and the Cultivation Project for FAST Scientific Payoff and Research Achievement of CAMS-CAS. T.H. acknowledges support from the European Research Council under the Horizon 2020 Framework Program via the ERC Advanced Grant Origins 83\,24\,28.

\end{acknowledgements}


\bibliographystyle{aa}
\bibliography{references}

\appendix

\section{Results with 1.1\,mm map included}
\label{app_withsma}

\begin{table}[htp]
\caption{Parameters in the best-fit model with 1.1\,mm included.}
\label{tab_model_app} \centering \small  
\setlength{\tabcolsep}{1.6mm}{
\begin{tabular}{lc|ccc}
\hline \hline
Parameter &  Unit & Range & Best fit & Uncertainties  \\
\hline
\multicolumn{2}{c}{Disk} \\
$r_{\rm in}$               & au     & $0.1\sim100.0$            & 16.0     & $^{+37.4}_{-8.0}$       \\
$r_{\rm out}$              & au     & $100.0\sim300.0$          & 176.0    & $^{+24.4}_{-19.1}$       \\
$h_{\rm out}/r_{\rm out}$  &        & $0.05\sim0.50$            & 0.22     & $^{+0.06}_{-0.06}$      \\
$p$                    &        & $0.5\sim2.0$                 & $0.83$   & $^{+0.34}_{-0.36}$      \\
$k$           &        & $1.0\sim5.0$                       & 1.165    & $^{+0.120}_{-0.180}$      \\
$\Sigma_{\rm out}$  & $\rm g\,cm^{-2}$  & $(0.1\sim10)\times10^{-2}$ & $4.9\times10^{-2}$ & $^{+1.0}_{-1.5}\times10^{-2}$ \\
$\phi$                     & deg    & $60.0\sim90.0$            & 88.0     & $^{+2.0}_{-5.2}$       \\
P.A.                       & deg    & $-20.0\sim-40.0$          & $-32.0$  & $^{-5.5}_{+5.8}$       \\
\multicolumn{2}{c}{Envelope} \\
$R_{\rm in}$               & au     & 0.1                       & 0.1      & Fixed     \\
$R_{\rm out}$              & au     & 900                       & 900      & Fixed     \\
$\rho_{\rm out}$    & $\rm g\,cm^{-3}$     & $(0.1\sim90)\times10^{-20}$  & $6.2\times10^{-20}$ & $^{+3.8}_{-2.3}\times10^{-20}$ \\
$\gamma$                   &        & $-0.01\sim-0.5$           & $-0.10$   & $^{-0.17}_{+0.10}$     \\
P.C.                       & deg    & 40.0                      & 40.0     & Fixed     \\
\multicolumn{2}{c}{Star} \\

$r_{\star}$                & \rsol   & 2.0                    & 2.0      & Fixed     \\
$M_{\star}$                & \msol   & 0.55                   & 0.55     & Fixed     \\
$T_{\star}$                & K      & $3000\sim5000$            & 4000     & $^{+248}_{-340}$        \\
$A_{\rm V}$                & mag    & $1.0\sim30.0$             & 12.91    & $^{+2.55}_{-1.46}$       \\
Distance                   & pc     & 140.0                     & 140.0    & Fixed     \\
\multicolumn{2}{c}{Dust opacity} \\
$a_{\rm max}^{\rm disk}$   & $\mu$m & $0.25\sim100\,000$        & 50\,000  & $^{+39\,000}_{-20\,400}$    \\
$a_{\rm max}^{\rm env}$    & $\mu$m & $0.25\sim100\,00$         & 110      & $^{+219}_{-69}$     \\
$a_{\rm min}^{\rm disk}$   & $\mu$m & 0.01                     & 0.01    & Fixed     \\
$a_{\rm min}^{\rm env}$    & $\mu$m & 0.01                     & 0.01    & Fixed     \\
\hline
\end{tabular}}
\begin{flushleft}
\textbf{Notes.} \\
$r_{\rm in}$: Inner radius of the disk.   \\
$r_{\rm out}$: Outer radius of the disk.   \\
$h_{\rm out}/r_{\rm out}$: Ratio of the pressure scale height to the radius at $r_{\rm out}$.   \\
$p$: Power exponent of the radial surface density distribution.   \\
$k$: Flare index.   \\
$\Sigma_{\rm out}$: Surface density of the disk at the outer radius.   \\
$\phi$: Inclination angle of the disk.   \\
$R_{\rm in}$: Inner radius of the envelope.   \\
$R_{\rm out}$: Outer radius of the envelope.   \\
$\rho_{\rm out}$: Density of the envelope at the outer radius.   \\
$\gamma$: Radial density distribution power exponent.   \\
P.C.: Radius of the polar cavity in the envelope.   \\
$r_{\star}$: Radius of the star.   \\
$M_{\star}$: Mass of the star, fitted with a Keplerian disk in the $^{12}$CO $J=2-1$ velocity field \citep{Launhardt2020}.  \\
$T_{\star}$: Effective surface temperature of the star.   \\
P.A.: The rotation of the major axis in the image plane.   \\
$A_{\rm V}$: Visual extinction. \\
Distance: The distance to the Sun.   \\
$a_{\rm max}^{\rm disk}$: Maximum dust grain size in the disk.   \\
$a_{\rm max}^{\rm env}$: Maximum dust grain size in the envelope.   \\
$a_{\rm min}^{\rm disk}$: Minimum dust grain size in the disk.   \\
$a_{\rm min}^{\rm env}$: Minimum dust grain size in the envelope. \\
Uncertainties: Defined by ${\chi}^2_{\rm total}-{\chi}^2_{\rm best}<3$.
\end{flushleft}
\end{table}

\begin{figure*}
\centering
\includegraphics[width=0.75\textwidth, angle=0]{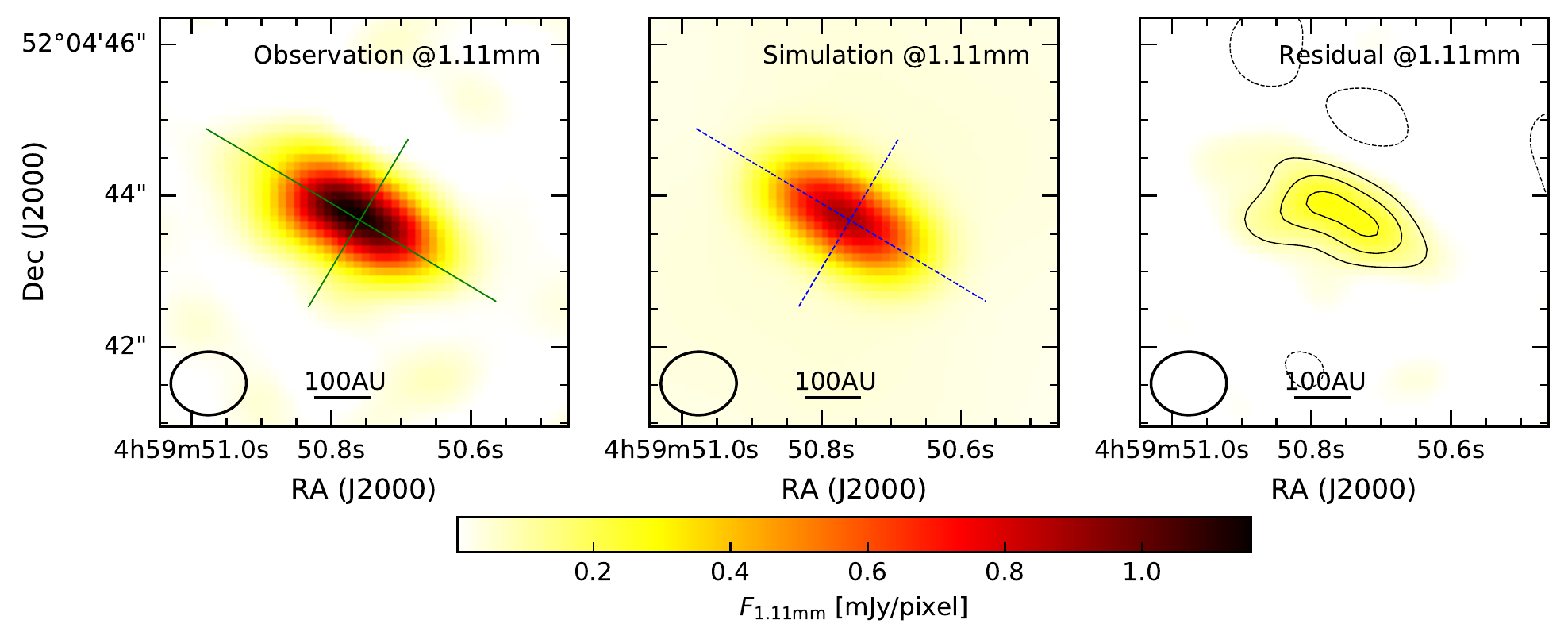}
\includegraphics[width=0.75\textwidth, angle=0]{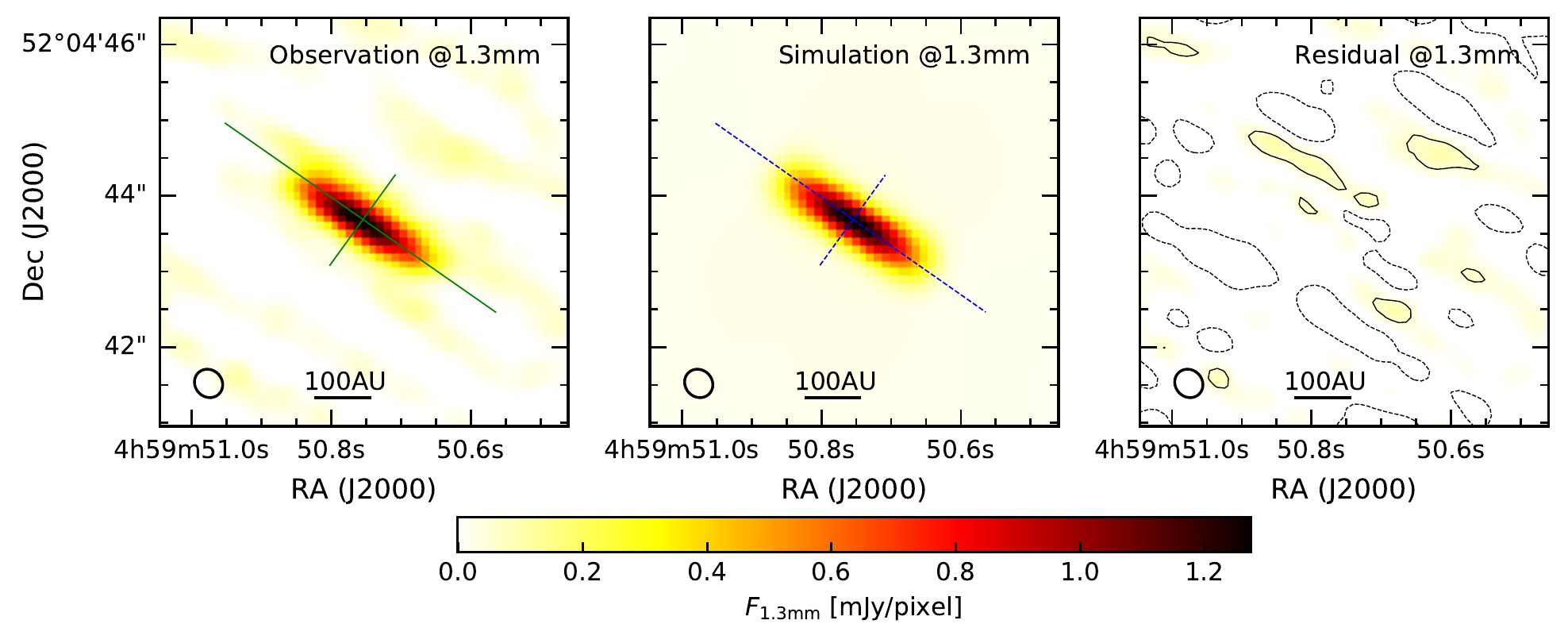}
\includegraphics[width=0.75\textwidth, angle=0]{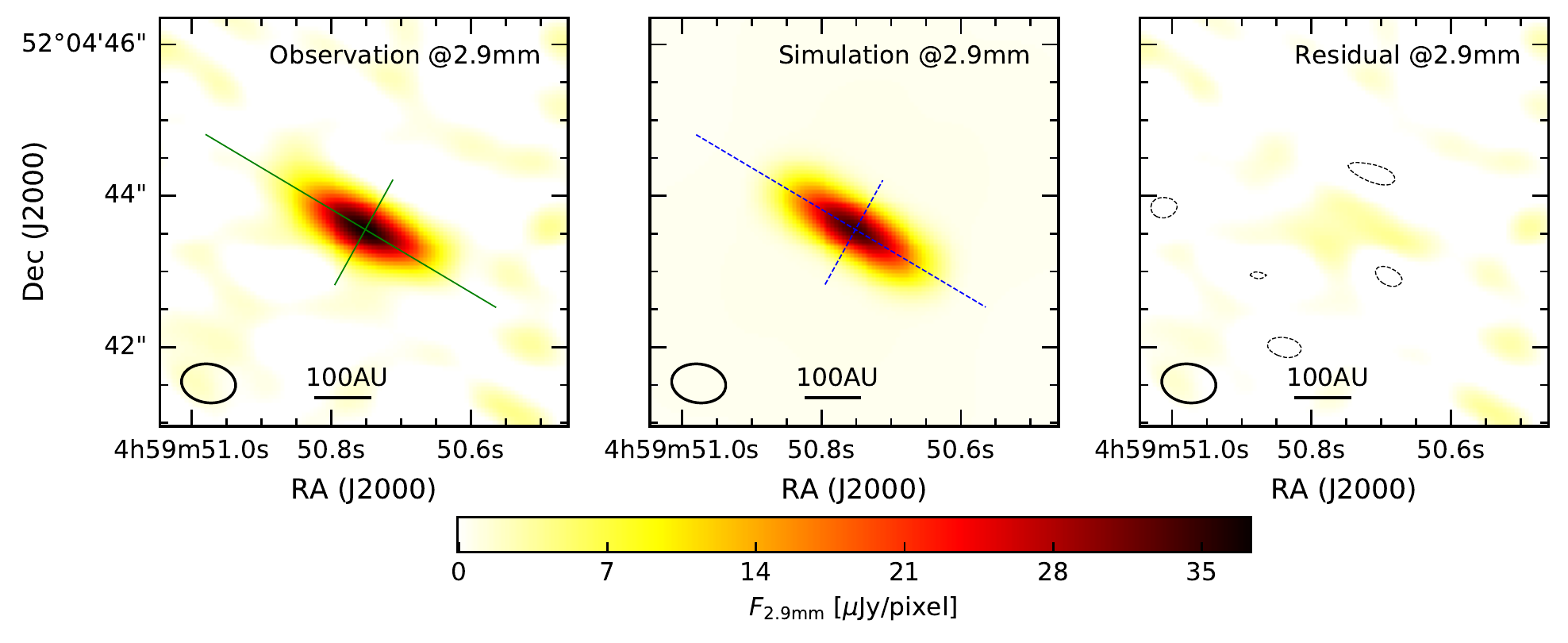}
\includegraphics[width=0.75\textwidth, angle=0]{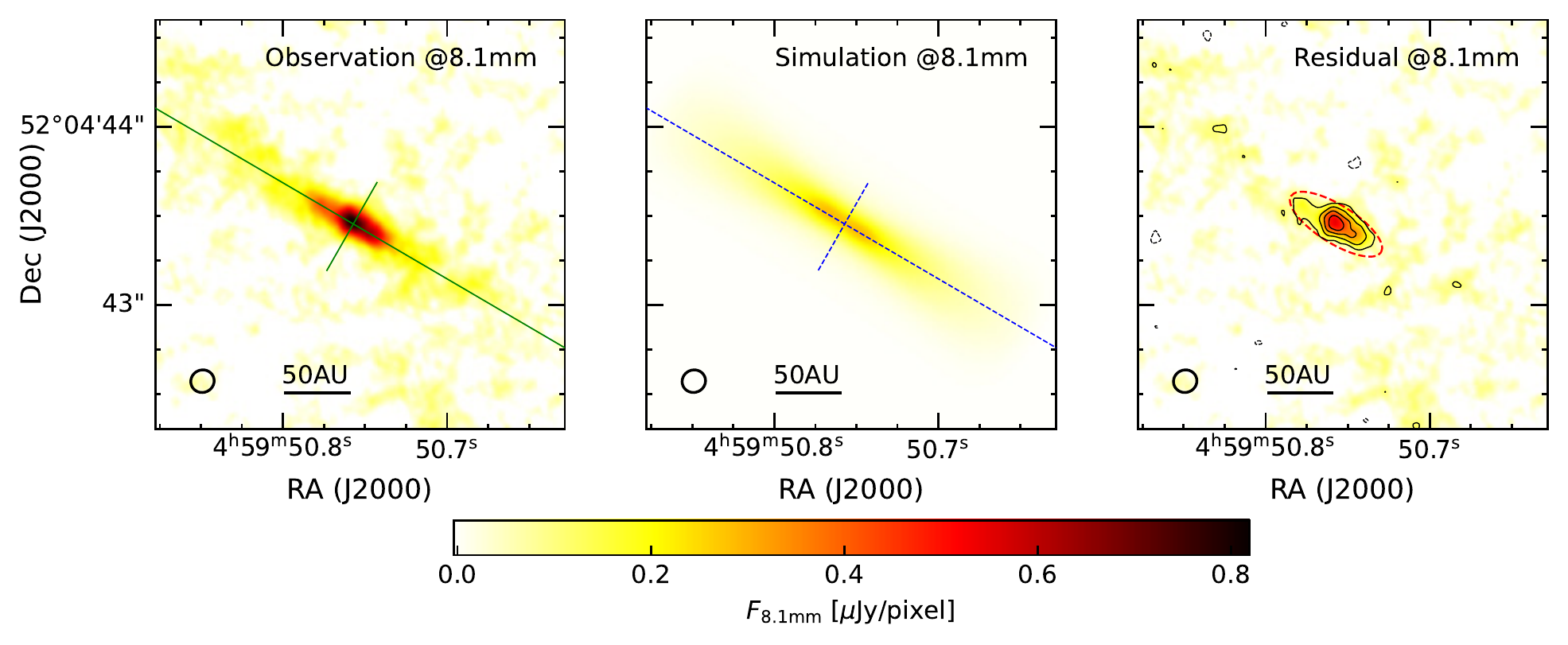}
\caption{Observed and modelled intensity maps at 1.1, 1.3, 2.9, and 8.1\,mm. The modelled maps have smaller pixel sizes than the observed maps, but are convolved to the same beam sizes as the respective observations (shown as ellipses in the lower left corners). The contour levels in each residual image start at -3$\sigma$, 3$\sigma$ in steps of 3$\sigma$ with $\sigma_{\rm 1.1mm}=0.028\,\mjyp$ ($\rm 1\,pixel=0.1''\times0.1''$ at 1.1\,mm), $\sigma_{\rm 1.3mm}=0.019\,\mjyp$ ($\rm 1\,pixel=0.1''\times0.1''$ at 1.3\,mm), $\sigma_{\rm 2.9mm}=1.37\,\mujyp$ ($\rm 1\,pixel=0.05''\times0.05''$ at 2.9\,mm), and $\sigma_{\rm 8.1mm}=0.038\,\mujyp$ ($\rm 1\,pixel=0.01''\times0.01''$ at 8.1\,mm) in the observations. Solid green and dotted blue lines show the directions of major and minor axes in the observation and simulation panels (cf. Fig.\,\ref{fig_obs_sim_1d_app}). The central compact emission within the red-dashed ellipse at 8.1\,mm residual map was masked for fitting (see details in Sect.\,\ref{sect_fit}).}
\label{fig_obs_sim_2d_app}
\end{figure*}

\begin{figure*}
\centering
\includegraphics[width=0.80\textwidth, angle=0]{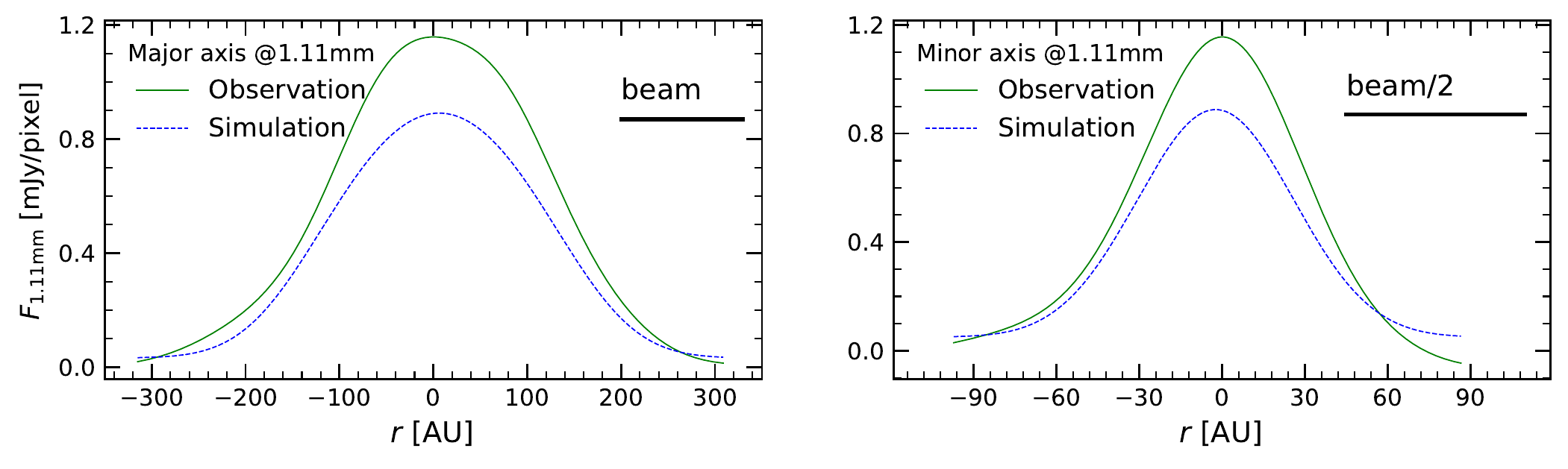}
\includegraphics[width=0.80\textwidth, angle=0]{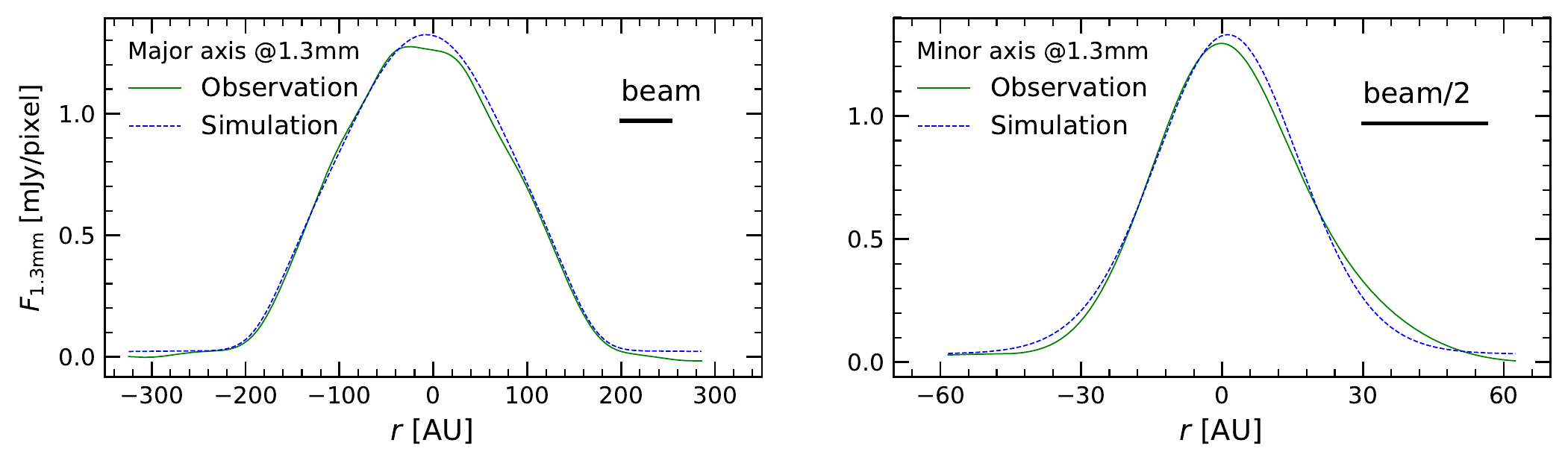}
\includegraphics[width=0.80\textwidth, angle=0]{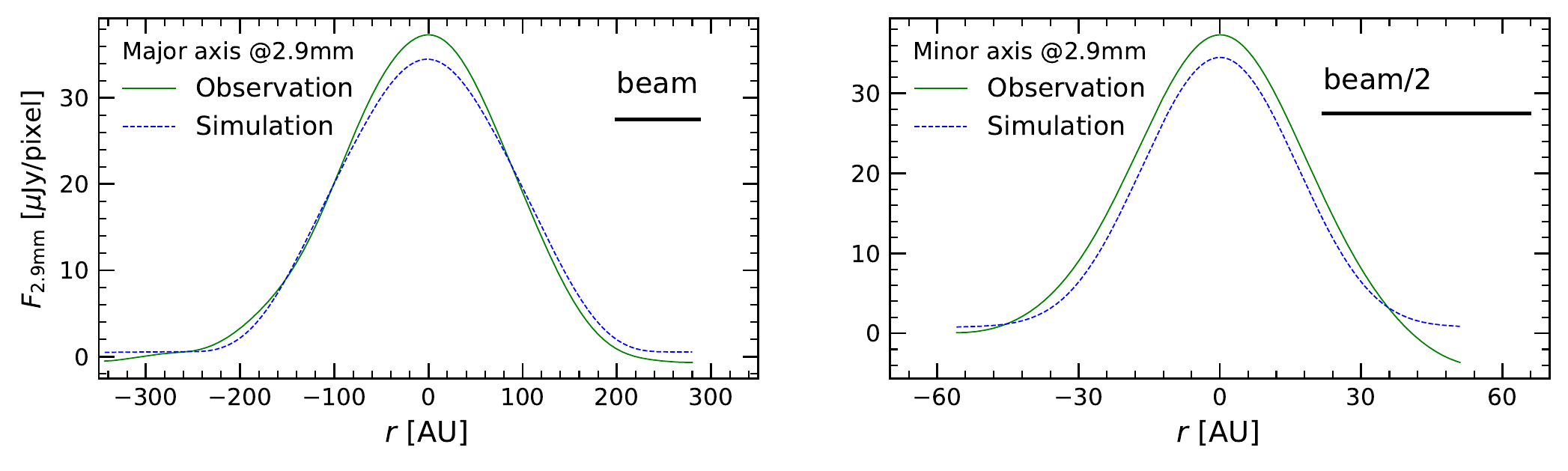}
\includegraphics[width=0.80\textwidth, angle=0]{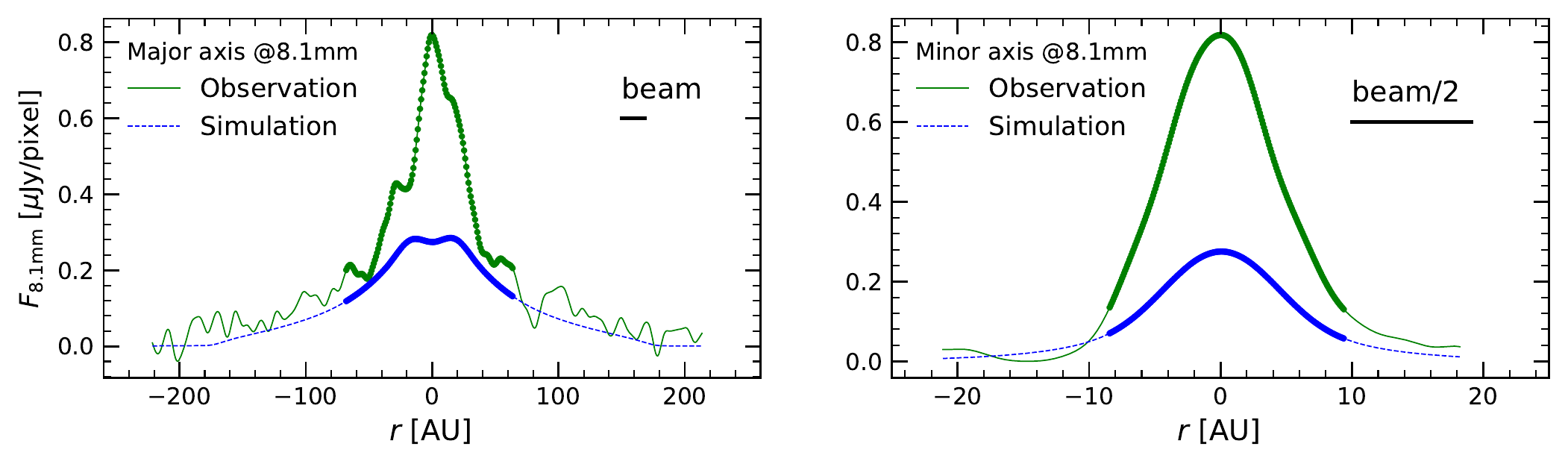}
\caption{Intensity distribution cuts along the major and minor axes (see Fig.\,\ref{fig_obs_sim_2d_app}) in observations and simulations at 1.1, 1.3, 2.9, and 8.1\,mm, respectively. The pixel size at each wavelength is listed in caption of Fig.\,\ref{fig_obs_sim_2d_app}. The central peak (marked with thick line) at 8.1\,mm was masked for fitting (see details in Sect.\,\ref{sect_fit} and Fig.\,\ref{fig_obs_sim_2d_app}).}
\label{fig_obs_sim_1d_app}
\end{figure*}

\clearpage

\label{lastpage}
\end{document}